\documentclass[fleqn,usenatbib]{mnras}

\usepackage{newtxtext,newtxmath}

\usepackage[T1]{fontenc}

\DeclareRobustCommand{\VAN}[3]{#2}
\let\VANthebibliography\thebibliography
\def\thebibliography{\DeclareRobustCommand{\VAN}[3]{##3}\VANthebibliography}

\usepackage{graphicx}	
\usepackage{amsmath}	
\usepackage{natbib}
\usepackage{textcomp}
\usepackage{bm}
\usepackage[normalem]{ulem}
\usepackage[cmyk]{xcolor}

\title[The Stellar "Snake" - II: The Mass Function]{The Stellar "Snake" - II: The Mass Function}

\author[Yang, et al.]{
Xiang-Ming Yang,$^{1,2}$
Sarah A. Bird,$^{1,2}$
Jiadong Li,$^{3}$
Hai-Jun Tian,$^{5,6}$\thanks{E-mail: hjtian@hdu.edu.cn, sarahbird@ctgu.edu.cn}
Dan Qiu,$^{4}$
Jia-Peng Li,$^{5,6}$
\newauthor Cheng-Yuan Li,$^{7,8}$
Gao-Chao Liu,$^{1,2}$
Peng Zhang,$^{1,2}$
Ju-Yong Zhang,$^{5,6}$
and Zhi-Ping Chen$^{5,6}$\\
$^{1}$Center for Astronomy and Space Sciences, China Three Gorges University, Yichang 443002, People's Republic of China\\
$^{2}$College of Science, China Three Gorges University, Yichang 443002, People's Republic of China\\
$^{3}$Max-Planck-Institut f{\"u}r Astronomie, K{\"o}nigstuhl 17, D-69117 Heidelberg, Germany\\
$^{4}$Key Lab of Space Astronomy and Technology, National Astronomical Observatories, Chinese Academy of Sciences,Beijing 100101, People's Republic of China\\
$^{5}$School of Science, Hangzhou Dianzi University, Hangzhou, 310018, People's Republic of China\\
$^{6}$Big Data Institute, Hangzhou Dianzi University, Hangzhou, 310018, People's Republic of China\\
$^{7}$School of Physics and Astronomy, Sun Yat-sen University, Zhuhai 519082, People's Republic of China\\
$^{8}$CSST Science Center for the Guangdong-Hong Kong-Macau Greater Bay Area, Zhuhai 519082, People's Republic of China\\
}

\date{Accepted XXX. Received YYY; in original form ZZZ}

\pubyear{2023}

\begin{document}
\label{firstpage}
\pagerange{\pageref{firstpage}--\pageref{lastpage}}
\maketitle

\begin{abstract}
We present a comprehensive investigation on the mass function (MF) of a snake-like stellar structure in the solar neighbourhood, building on our previous discovery. 
To ensure the reliability of the data, we reselect the member stars of the Stellar ``Snake'' in the latest {\it Gaia} Data Release 3 using the same approach as the initial series of articles. We also precisely measure the physical parameters of the clusters within the Stellar Snake. In light of the high completeness of the member stars in the cluster regions, we develop a simulated model color-magnitude diagram-based inference method to derive the mass function, binary fraction, and mass-ratio distribution of the clusters in the Stellar Snake.
Notably, despite their similar ages and metallicity, we discover systematic variations in the MFs along the elongation direction of the Snake in the mass range of 0.5 to 2.0 M$_\odot$. 
The ``head'' of the Snake conforms to a canonical initial mass function with a power-law slope of $\alpha\sim-2.3$. 
Extending towards the ``tail,'' the MF becomes more top-light,
indicating a deficiency of massive stars within these clusters.
This result provides evidence for the delayed formation of massive stars in the clusters. Such clues give support to the hypothesis that the Stellar Snake constitutes as a hierarchically primordial structure.

\end{abstract}




\section{Introduction}

The stellar initial mass function (IMF) describes the distribution of the number of stars formed as a function of their mass. 
Originally proposed by \cite{salpeter1955luminosity} and further studied by seminal works like \cite{miller1979initial} and \cite{kroupa2001variation}, the IMF is typically described as a power-law distribution. 
At high masses, the IMF follows a power-law slope of around $-2.3$, while slopes of around $-1.3$ and $-0.3$ are seen at intermediate and low masses, respectively.

The IMF is fundamentally important because it determines the initial properties of stellar populations and galaxies as they evolve over time. 
In particular, the shape of the IMF at its low-mass end, encompassing brown dwarfs, very-low-mass stars and low-mass stars, has crucial implications for understanding the unseen, or ``dark", mass budget of galaxies.
Furthermore, a well-determined IMF
is an essential ingredient to
many astrophysical models, including galaxy dynamics and stellar population synthesis; such models rely on assumptions and observations related to the IMF \citep{romano2006formation,lamers2013evolution}. 
However, observational limitations pose challenges to measuring the IMF. 
Even within a few kpc of the Sun, we can only completely survey the most massive stars, and low-mass stars remain elusive beyond such volume.
Moreover, direct mass measurements are only feasible for binary systems, yet binary orbits are unavailable for the vast majority of stars. 
As a result, masses must typically be inferred through luminosities, colors, and spectra compared to theoretical stellar models.


Earlier works assumed a universal IMF \citep[canonical IMF,][]{kroupa2001variation} based on studies finding its robustness across many environments \citep[e.g., ][]{bastian2010universal}.
However, highly improved observations now probing a diversity of astrophysical environments and physical scales are revealing the IMF may not be constant, as recent evidence of IMF variations demonstrates. 
\cite{bastian2010universal} and \cite{zhang2018stellar} find top-heavy IMFs measured in high star formation rate densities like galactic centers and starburst galaxies. 
\cite{van2010substantial} and \cite{conroy2012stellar} find bottom-heavy IMFs detected in metal-rich environments. \cite{geha2013stellar} measure bottom-light IMFs observed in metal-poor Milky Way dwarf satellites.

Further observational constraints on the IMF under a more full range of astrophysical conditions are needed to elucidate what drives its diversity and properly account for IMF variations in galaxy models. Specifically, observational studies have shown that stellar populations forming at early cosmic times contained a deficit of low-mass stars compared to predictions of the canonical IMF, and the present-day IMF exhibits an enhanced proportion of low-mass stars at higher stellar metallicities, as demonstrated through large-scale spectroscopic surveys \citep{2023Natur.613..460L}. Understanding IMF variations remains an important outstanding problem with wide-ranging implications.

Achieving a robust characterization of the IMF requires deeper insight into stellar birth and evolution.
Theoretical models predict the IMF arises from fragmentation and competitive accretion within clustered environments, implying dependencies on the initial composition and structure of natal clouds \citep{jappsen2005stellar,chabrier2014variations}.
Apart from direct measurements of the present-day IMF from the prestellar and star forming natal clouds, studying the present-day MF (PDMF) of young stellar systems is a close avenue for gaining insight on the present-day IMF.

Open clusters and associations are excellent sites to measure the 
PDMF. This function is defined as the relative proportion of coeval stars with different masses. This quantity reflects the initial physical conditions (density and metal content) affecting the fragmentation efficiency of the primordial cloud from which these stars form \citep{hennebelle2008analytical}. Open clusters are composed of coeval and chemically homogeneous stars located at the same distance, thus eliminating the uncertainties linked to relative distance and differences in the age and chemical composition of individual stars \citep{ebrahimi2022family}. Consequently, this allow us to avoid potential influences from different stages of evolution and metallicity trends.

In recent years, numerous studies have focused on determining the PDMF of open clusters \citep[e.g.,][]{massey1995initial,slesnick2002star,moraux2004pleiades,pandey2007stellar,dib2014testing,kalari2018magellanic}. Due to various factors such as dynamical evolution, differences in metallicity, variations in star-forming environments, and the relatively short time needed
for massive stars to leave the main sequence, the MFs of these open clusters do not always perfectly match the distribution expected for the canonical IMF. 

In the case of some older open clusters, dynamical evolution has led to mass segregation, where more massive stars are concentrated toward the center of the cluster, while less massive stars have gradually moved outward beyond the cluster's radius \citep{schneider1979catalog,portegies2010young}. This results in the PDMF of some older star clusters becoming flatter, such that the proportion of high-mass stars relative to the canonical IMF is higher \citep{kraus2007stellar, goldman2013towards, bisht2020comprehensive}. 
\textbf{\cite{ebrahimi2022family} also uncover a similar phenomenon. They find that open clusters with ages less than half of the half-mass relaxation time exhibit an MF resembling the canonical IMF. On the other hand, for clusters with ages equal to or greater than half of the half-mass relaxation time, the MF tends to become flatter.}
\textbf{However, in a study of 773 open clusters, \cite{almeida2023revisiting}  find no clear trend in age-related mass segregation.}
\textbf{In general, the majority of open clusters exhibit MFs that are similar to the canonical IMF \citep{bonatto2005spatial, niedzielski2007planetary, angelo2019characterizing, cordoni2023photom}.}

For relatively young open clusters and associations, the PDMF is more akin to the IMF because the majority of member stars are located in the main-sequence or pre-main sequence stages; they have yet to evolve long enough for the massive stars to leave the main sequence. Additionally, the impact of dynamical evolution is not as pronounced as in older clusters. However, the number of cluster members will decrease with the dynamic evolution of the cluster, 
even within a few tens of millions of years from its formation \citep{dinnbier2022majority}. Consequently, for these relatively young clusters, while they are less affected by dynamical evolution and the departure of high-mass stars from the main sequence,
the PDMF is slightly different from the IMF.
In spite of this, the PDMF remains crucial for uncovering the patterns of star formation and temporal variations in dynamic processes within young clusters, and exposing the physical characteristics of star-forming regions. Such research not only contributes to a comprehensive understanding of the evolutionary processes of star clusters but also provides essential clues for further exploration of the characteristics and formation mechanisms of the IMF.

The Stellar Snake \cite[first reported by][]{tian2020discovery} is a more recently discovered filamentary structure composed of at least nine open clusters \citep{wang2022stellar}. 
This formation mechanism of the Stellar Snake was elucidated by the identification of coeval stars that bridge the open clusters, as demonstrated by \citet{beccari2020uncovering}.
The Stellar Snake consists of numerous member stars with approximately the same age ($30-40$\,Myr) and metallicity (solar metallicity), and the structure is located at a relatively close distance (300 pc from the Sun). Under the conditions of relatively fixed metallicity and age (ranging only from $30-40$\,Myr), the Snake is likely not significantly dynamically evolved and most member stars are likely retained. The Snake is expected to exhibit a relatively uniform MF that closely resembles the IMF. Therefore, the Snake provides an ideal laboratory for analyzing stellar formation and evolution in cluster environments as well as tracking the IMF.

On top of the Snake constituting as an ideal laboratory, $\it{Gaia}$ DR3 \citep{vallenari2023gaia} presents unprecedentedly accurate and precise information for over 1.59 billion sources. These data include positions ($l$,\,$b$), proper motions ($\mu_l$,\,$\mu_b$), and parallaxes $(\varpi)$ with average systematic uncertainties ranging from 0.05$-$0.1 mas (for $G < 20$\,mag). Additionally, high-precision measurements are available for $G$, $G_\mathrm{BP}$, and $G_\mathrm{RP}$ magnitudes, with typical errors ranging from 0.3$-$6.0 mmag (for $G < 20$\,mag). These advancements enable us to make more accurate determinations of the members of the star clusters within the Stellar Snake, consequently, obtain more precise parameters and more reliable MFs for each cluster.

In this paper, we investigate the variations in the PDMF of the open clusters in the Stellar Snake. To do this we utilize data from {\it Gaia} DR3 and construct a simulated CMD model to calculate the PDMF for nine member clusters of the Snake.
In Section\,\ref{sec:data}, we utilize data from $\it{Gaia}$ DR3  to identify member stars of the open clusters within the Stellar Snake.
In Section\,\ref{sec:model}, we construct a robust simulated CMD model to calculate the cluster's MF ($\alpha$), binary fraction ($f_b$), and binary mass ratio distribution ($\gamma$). In Section\,\ref{sec:result}, we present the results characterizing the Stellar Snake open clusters using our simulated CMD model; we also present intriguing trends in the spatial distribution along the Snake of the cluster MFs and give a potential physical explanation of our results. In Section\,\ref{sec:discussion}, we evaluate the selection of the open cluster member stars and the performance of our model, as well as compare the MF derived from the Stellar Snake field stars to those of the open clusters. Finally, we conclude in Section\,\ref{sec:conclusions}.

\section{Data}\label{sec:data}
In this section, we carefully select the member stars of the Stellar Snake (Section \ref{Snake members})  and its open clusters as well as derive essential cluster parameters (Section \ref{Asteca}), such as stellar age, metallicity, cluster radius ($r_\mathrm{cl}$) and spatial distribution. By leveraging off of the comprehensive {\it Gaia} DR3 dataset and implementing a systematic approach, we ensure high quality and quantity samples that contribute to a comprehensive understanding of the Snake's characteristics. Considering the high demand for a complete sample from which to build the MF, the overall MF of the Snake (field stars and cluster members) may be affected by incompleteness. Therefore, we primarily focus on studying the open clusters within the Snake to enhance the reliability of our results. Based on the member stars of the Stellar Snake data and using the parameters specific to the star clusters, we identify a reliable sample of member stars for each cluster (Section \ref{subsection:final_members}). Finally, we validate the completeness of the sample to ensure a more accurate reflection of the true properties of the clusters (Section \ref{completeness}).

\subsection{Selection of Stellar Snake Member Stars}\label{Snake members}

Firstly, we select member stars of the Snake by 5D phase information (i.e.$, l, b, \mu_{l*} , \mu_{b},$ and
distance). We adopt the solar motion ($U_\odot$, $V_\odot$, $W_\odot$) = (9.58, 10.52, 7.01) km s$^{-1}$ \citep{tian2015stellar} with respect to the local standard of rest, and the solar Galactocentric radius and vertical height ($R_0, Z_0$) = (8.27, 0.0) kpc \citep{schonrich2012galactic}. In the gnomonic projection coordinate system, we use $l^*$ to denote the Galactic longitude, such that, for example, $\mu_l* = \mu_l\cos{b}$. 
We correct the proper motions ($\mu_l* , \mu_b$) of each star for the solar peculiar motion.

In order to select member stars of the Stellar Snake, we first need to impose certain restrictions on the search range and filtering criteria to obtain preliminary samples.
Based on the previously discovered adjacent serpent-like \citep{tian2020discovery} and filamentary \citep{beccari2020uncovering} structures, two search regions, Part I and Part II, are defined to study the interconnection between these structures.
We follow the empirical criteria mentioned by \cite{wang2022stellar} and re-screen the samples from Parts I and II using {\it Gaia} DR3.

\begin{figure}
\centering
\includegraphics[width=1.0\linewidth]{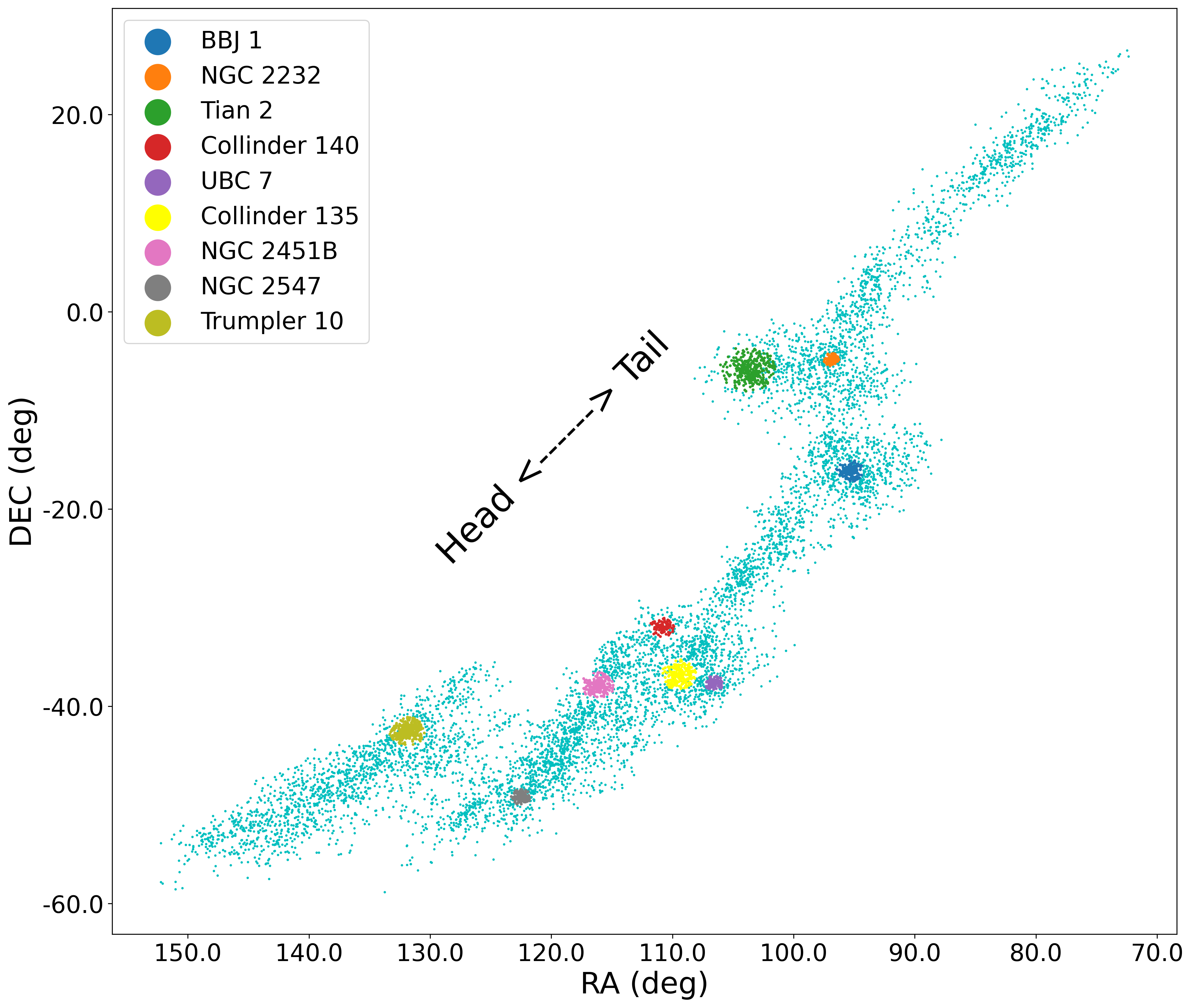}
\caption{\label{fig:radeclb}Spatial distribution of candidate member stars in the Stellar Snake in terms of (RA, DEC). The cyan background points represent the member field stars of the filamentary Stellar Snake structure (Section \ref{Snake members}), our curated samples of member stars for each selected open clusters, detailed in Section \ref{subsection:final_members}, is highlighted with distinct colors corresponding to each cluster as indicated in the legend. 
The ``head'' to the ``tail'' of the structure is annotated for orientation.
\label{radec}}
\end{figure}

After data selection, we adopt the FoF algorithm using the software ROCKSTAR \citep{behroozi2012rockstar} to search for members of the Snake from both Parts I and II. FoF is an algorithm used to search for members of a group; this algorithm has been well implemented in ROCKSTAR \citep{behroozi2012rockstar}, which employs a technique of adaptive hierarchical refinement in 6D phase space to divide all stars into several FoF groups by tracking the high number density clusters and excluding those stars that could not be grouped in star aggregates.  
Taking into account the completeness of the $\it{Gaia}$ data, the member stars might not be complete at fainter magnitudes, potentially impacting the results of the MFs. We conservatively impose a limit on the $G$-band magnitude to be less than 18\,mag for our subsequent research in order to ensure the completeness of the sample. Finally we retain 2217 and 6991 candidate Snake member stars of Parts I and II, respectively.
In Figure \ref{fig:radeclb}, the cyan markers illustrate the distribution of the Snake members with $G < 18$\,mag in coordinates (RA, DEC). 
For ease of description, we spatially categorized the Snake based on its distribution \citep{wang2022stellar}. The region with relatively older ages, specifically Trumpler\,10 (high $RA$), is defined as the ``head" of the Snake, while the region with relatively younger ages, Tian\,2 (low $RA$), is defined as the ``tail" of the Snake.

\subsection{Parameter Determination of Open Clusters in the Stellar ``Snake''}\label{Asteca}

In order to study the mass function of each cluster in the Snake, it is essential to determine the fundamental cluster parameters accurately. 
This includes utilizing precise isochrones characterized by metallicity and age to establish reliable evolutionary tracks for the member stars of the clusters in the CMD. 
Additionally, considering the distance of the clusters from us, we need to factor in the distance modulus to convert apparent magnitudes to absolute magnitudes. On the other hand, extinction makes stars appear redder and dimmer, requiring corrections to restore their true luminosity. Therefore, accounting for the effects of the distance modulus and extinction is crucial for accurately describing the distribution of stars of different masses in the CMD.
These considerations enable us to determine the masses of member stars and gain insights into the evolutionary processes of the clusters.

We obtain PARSEC\footnote{\label{link} \href{http://stev.oapd.inaf.it/cgi-bin/cmd}{http://stev.oapd.inaf.it/cgi-bin/cmd}} \citep{bressan2012parsec} isochrones ranging from 0.0005 to 0.0305 in metallicity (Z) and from 6 to 9 in logarithmic age (log Age), incrementing by 0.0005 and 0.005, respectively. 
We utilize the package {\tt Automated Stellar Cluster Analysis} 
\citep[{\tt ASteCA,}][]{perren2015asteca} to determine the cluster parameters. 
The process of determining the cluster center and radius using {\tt ASteCA} involves several steps. Below we summarize the steps taken by {\tt ASteCA} to select the initial cluster members:

\begin{enumerate}
    \item Input Nearby Samples: {\tt ASteCA} takes as input a sample of stars that are near the suspected star cluster. This sample typically includes both cluster members and field stars. Taking into consideration that {\tt ASteCA} can only analyze one star cluster at a time, for each star cluster, we select a preliminary sample from the data obtained from Section\,\ref{Snake members}. This preliminary sample includes stars about the approximate central region within a radius of 3 to 5 degrees and with $G < 18$\,mag as input.

    \item 
     Fitting the Radial Density Profile (RDP): {\tt ASteCA} calculates the RDP of the input sample. 
    The RDP represents the spatial distribution of stars as a function of distance from the cluster center, providing information about the cluster's extent.

    \item Cluster Radius ($r_\mathrm{cl}$) Determination: Using the RDP, {\tt ASteCA} determines the $r_\mathrm{cl}$ by identifying the point where the density of stars transitions from the higher density of the cluster region to the lower density of the field region. This transition point corresponds to the estimated radius of the star cluster, which marks the estimated cluster boundary. 

    \item Initial Sample Mixing: {\tt ASteCA} creates an initial sample by mixing the input sample of suspected cluster members with the nearby field stars. This mixed sample helps establish a more robust statistical analysis.

    \item Fitting PARSEC Isochrones: {\tt ASteCA} employs the Markov Chain Monte Carlo (MCMC) method to fit the mixed sample with the corresponding PARSEC isochrones. 
The MCMC analysis 
infer the parameters of age ($\log\mathrm{Age}$), metallicity ($Z$), extinction ($A_V$, where $A_G = A_V\times 0.836$, $A_\mathrm{BP\_RP}= A_V\times 0.449$), and distance modulus ($\mathrm{DM}=(m-M)_0$) to find the best-fitting values that reproduce the observed properties of the stars in the mixed sample.

    \item Determining Cluster Parameters: Through the MCMC posteriors, {\tt ASteCA} determines the best-fitting cluster parameters, including age, metallicity, extinction, and distance modulus. These parameters provide insights into the characteristics and evolutionary stage of the star cluster.
    
\end{enumerate}

By following these steps, {\tt ASteCA} facilitates the determination of the cluster center, $r_\mathrm{cl}$, and other important parameters, enabling a comprehensive analysis of the star clusters based on their observed properties and theoretical isochrones.

\begin{figure}
\centering
\includegraphics[width=1.0\linewidth]{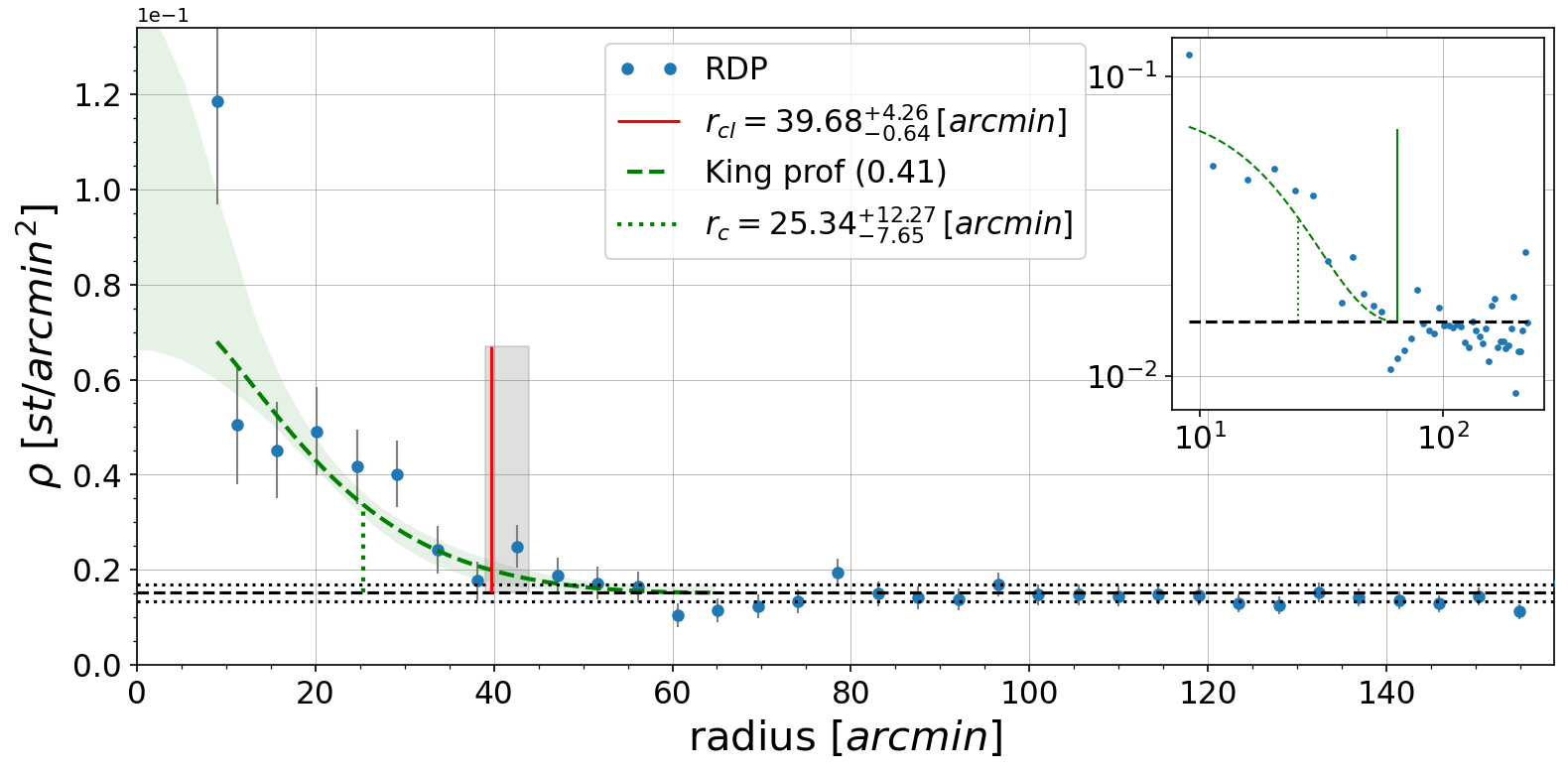}
\caption{\label{fig:NGC2232_A}Radial density profile (RDP) of the open cluster NGC\,2232 derived using {\tt ASteCA}. Blue points represent the stars/arcmin$^2$ RDP with the cluster center as the origin; the corresponding error bars are the standard deviations. The horizontal black dashed line is the field density value
and the black dotted lines represent its upper and lower limits. The red vertical line is the derived cluster radius $r_\mathrm{cl}$ with the uncertainty region marked as the gray shaded area. A 2P King profile fit is indicated with the green dashed curve and its associated uncertainties are depicted by the green shaded region with the core radius $r_\mathrm{c}$ shown as a green vertical dotted line.
The inset plot in the upper right corner displays the results in logarithmic coordinates.}
\label{para_ngc2232}
\end{figure}

We employ {\tt ASteCA} to obtain the best-fit parameters for the open clusters in the Stellar Snake. 
Here we show the results for the cluster NGC\,2232 as an working example. 
Figure \ref{fig:NGC2232_A} shows the respective radii and associated uncertainties of the star cluster's RDP, and Figure \ref{fig:NGC2232_B} shows the MCMC results for the optimal age, metallicity, distance modulus, and extinction of the star cluster using the best-fitting PARSEC isochrones. 


Finally, we obtain the fundamental parameters for the nine open clusters in the Stellar Snake through {\tt ASteCA} and summarize the results in Table \ref{tab:Parameters}. 

\begin{table*}
\centering
\caption{\label{tab:Parameters}Derived Stellar Snake open cluster parameters.}
\scalebox{0.82}{
\begin{tabular}{c|c|c|c|c|c|c|c|c|c|c|c|c}
\hline
Cluster & $N$ & $N^*$ & RA$^a$ & DEC$^a$ & ${\mu^*_\mathrm{RA}}^a$ & ${\mu_\mathrm{DEC}}^a$ & ${\varpi}^a$ & ${\log{\mathrm{Age}}}^a$ & ${A_V}^a$ & ${Z}^a$ & DM$^a$ & ${r_\mathrm{cl}}^a$  \\\hline
 & & & \multicolumn{2}{|c|}{deg} & \multicolumn{2}{c|}{mas/yr} & mas & dex & mag &  & mag & arcmin\\\hline
 BBJ\,1 & 193 & 228 & 95.36825 & -16.09487 & -5.321 & -4.826 & $2.750^{+0.002}_{-0.003}$ & $7.576\pm0.074$ & $0.318\pm0.145$ & $0.0184\pm0.0018$ & $7.99\pm0.12$ & $67.10^{+6.03}_{-1.72}$\\\hline
NGC\,2232 & 134 & 147 & 96.91900 & -4.80574 & -4.738 & -1.701 & $3.166^{+0.003}_{-0.003}$ & $7.499\pm0.072$ & $0.250\pm0.088$ & $0.0149\pm0.0014$ & $7.41\pm0.13$ & $39.68^{+4.26}_{-0.64}$\\\hline
Tian\,2$^b$ & 339 & 371 & 103.83076 & -5.72400 & -7.401 & -2.426 & $3.525^{+0.002}_{-0.003}$ & $7.493\pm0.062$ & $0.199\pm0.107$ & $0.0165\pm0.0013$ & $7.20\pm0.13$ & $139.40^{+11.00}_{-1.84}$\\\hline
Collinder\,140 & 117 & 129 & 110.85415 & -31.92475 & -7.900 & 4.649 & $2.613^{+0.002}_{-0.002}$ & $7.568\pm0.093$ & $0.220\pm0.096$ & $0.0200\pm0.0035$ & $8.04\pm0.17$ & $61.16^{+6.50}_{-0.42}$\\\hline
UBC\,7 & 104 & 119 & 106.55510 & -37.64421 & -9.611 & 7.082 & $3.663^{+0.002}_{-0.003}$ & $7.552\pm0.132$ & $0.259\pm0.186$ & $0.0166\pm0.0024$ & $7.26\pm0.21$ & $46.25^{+1.68}_{-1.89}$\\\hline
Collinder\,135 & 224  & 262 & 109.53068 & -36.76387 & -10.124 & 6.145 & $3.374^{+0.002}_{-0.001}$ & $7.606\pm0.066$ & $0.180\pm0.064$ & $0.0174\pm0.0034$ & $7.51\pm0.12$ & $90.66^{+1.64}_{-16.97}$\\\hline
NGC\,2451B & 270 & 306 & 116.19883 & -37.90120 & -9.523 & 4.792 & $2.771^{+0.001}_{-0.001}$ & $7.501\pm0.054$ & $0.474\pm0.072$ & $0.0211\pm0.0012$ & $8.09\pm0.11$ & $81.00^{+1.93}_{-17.35}$\\\hline
NGC\,2547 & 241 & 277 & 122.51501 & -49.18754 & -8.573 & 4.477 & $2.626^{+0.002}_{-0.001}$ & $7.499\pm0.059$ & $0.308\pm0.078$ & $0.0186\pm0.0015$ & $8.12\pm0.10$ & $48.49^{+1.93}_{-6.88}$\\\hline
Trumpler\,10 & 364  & 435 & 131.89938 & -42.49871 & -12.382 & 6.652 & $2.332^{+0.001}_{-0.001}$ & $7.596\pm0.030$ & $0.278\pm0.057$ & $0.0215\pm0.0019$ & $8.42\pm0.06$ & $95.50^{+9.98}_{-14.78}$\\\hline
\end{tabular}}
\begin{minipage}{\textwidth}
\raggedright
\textbf{Note: } $N$: Final number of cluster member stars after applying the criteria RUWE $< 1.4$. (Section\,\ref{subsection:final_members}).\\
$N^*$: Final number of cluster member stars without applying the RUWE restriction. \\
$^a$Parameters obtained using {\tt ASteCA} (Section \ref{Asteca}).\\
$^b$Also called LP\,2439 in \cite{pang2020different}.
\end{minipage}
\end{table*}

\subsection{Member Stars of Open Clusters}
\label{subsection:final_members}
Based on the star cluster radius ($r_\mathrm{cl}$) and cluster centers obtained from {\tt ASteCA}, 
we select the corresponding RA and DEC coordinates from Table \ref{tab:Parameters} as the centers of each cluster.
We select all Stellar Snake member stars within $r_\mathrm{cl}$ as the 
final sample of member stars of each open cluster.

Subsequently, we calculate the median parallax value and its associated error for all candidate stars within the selected radius range. {\it Gaia} data provides precise parallax measurements, allowing us to obtain reliable samples of member stars for all open clusters in the Stellar Snake by restricting the parallax values. Applying excessively strict selection criteria may lead to the exclusion of a significant number of member stars, resulting in potential biases in the results. Therefore, we apply the parallax restriction only to the data within the cluster's radius to strike a balance between sample reliability and inclusivity. We calculate the median parallax and its standard deviation for all candidate stars within $r_\mathrm{cl}$, and retain candidate stars whose parallax is within the range of the median value $\pm 3$ times the corresponding error (within a $3\sigma$ range). The outliers are either spurious members, or true Snake members but not cluster members.

These selected stars are considered as the member stars of the open clusters. The final number of member stars $N$ for each star cluster is shown in Table \ref{tab:Parameters}. Figure \ref{fig:radeclb} shows the distribution of our final Snake cluster member star samples in coordinates (RA, DEC). 


\subsection{Completeness Verification}\label{completeness}

In calculating the MF, balance is essential between the quality of the data and the completeness of the sample. The MF is primarily influenced by two factors: on one hand, the completeness of {\it Gaia} itself, which unavoidably results in varying completeness under different conditions and in different regions due to the observational limitations of {\it Gaia}; on the other hand, our quality control measures for the initial sample selection. We apply criteria such as RUWE $< 1.4$ (RUWE is a normalized chi-square value provided in the data releases that is obtained by fitting {\it Gaia} sources to the point-spread function) and $\varpi/\delta_{\varpi} > 10.0$ to choose a higher quality sample. These conditions, to some extent, have implications for data completeness.

To investigate the first aforementioned factor that influences the MF, we employ the empirically built analytical model\footnote{\label{link2} available online as part of the {\tt gaiaunlimited} Python package at \href{https://github.com/gaia-unlimited/gaiaunlimited}{https://github.com/gaia-unlimited/gaiaunlimited}} proposed by \citet{cantat2023empirical}
to test the completeness of all the {\it Gaia} data within the {\tt ASteCA}-derived open cluster $r_\mathrm{cl}$ centered on the cluster centers (Table \ref{tab:Parameters}).
The model is calibrated using the deep imaging of the Dark Energy Camera Plane Survey {\it DECaPS} as a complete sample and predicts {\it Gaia}’s completeness values with a precision of up to a few per cent across the sky.
To develop their model, \citet{cantat2023empirical} explore several quantities and find that the best indicator of completeness is the $G$ magnitude of the sources with the smallest number of observations.
They make use of the number of observations used to compute the astrometric solution of a given source which is labeled in the {\it Gaia} DR3 catalogue as {\tt astrometric\_matched\_transits}.
Defining $M_{10}$ as the median magnitude of {\it Gaia} sources within a patch of sky with {\tt astrometric\textunderscore matched\textunderscore transits} $\leq 10$, \citet{cantat2023empirical} build their model to depend on $M_{10}$ as a proxy for completeness and find the model accounts for both the effects of crowding and the {\it Gaia} scanning law. We use their online package to assess the completeness of our data at various $G$ magnitudes within the region of each open cluster.

\begin{figure*}
\centering
\includegraphics[width=1.0\linewidth]{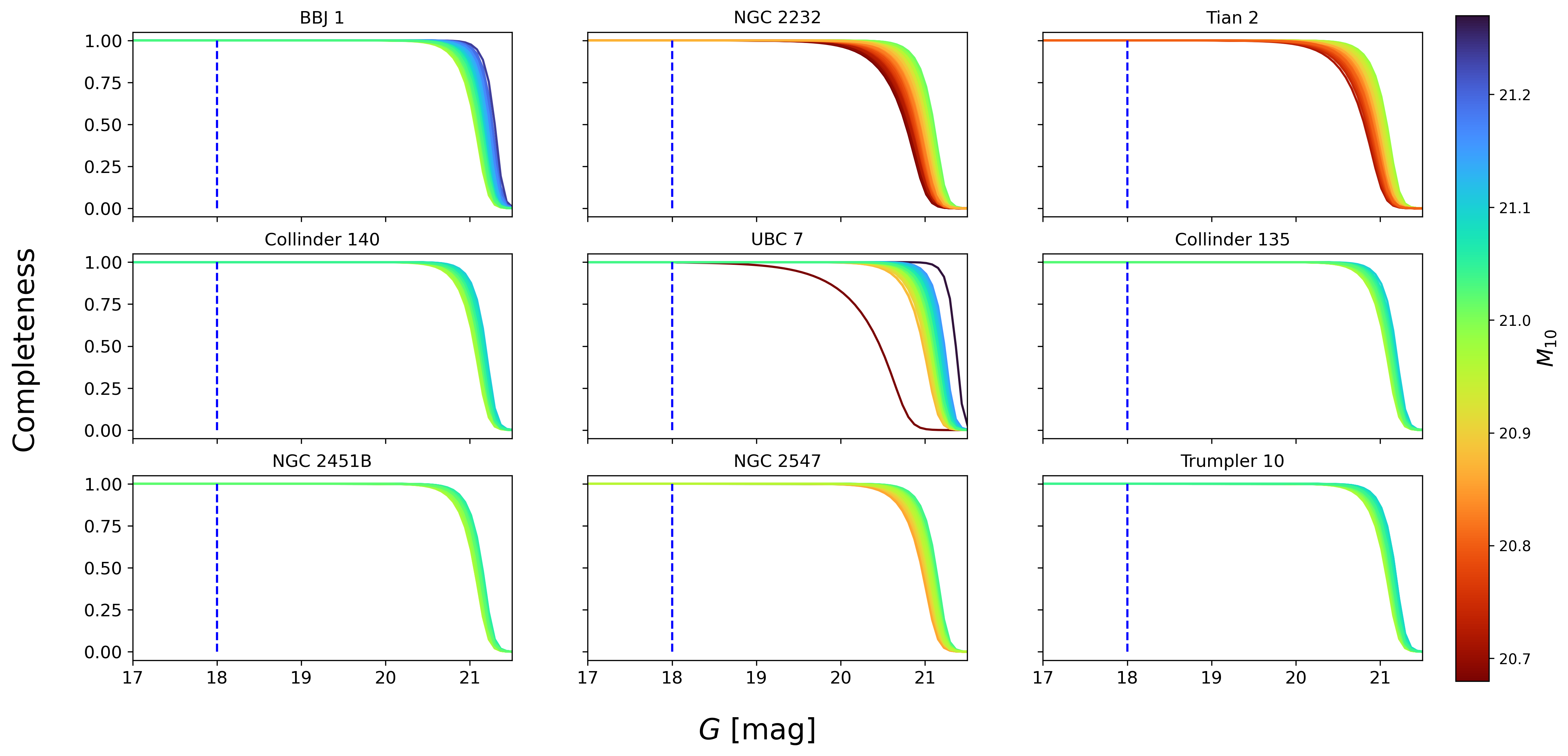}
\caption{\label{fig:completeness_cluster}Completeness analysis of all {\it Gaia} data sources  within  $r_\mathrm{cl}$ for the nine Stellar Snake open clusters. The $x$-axis represents the $G$-band magnitude, the $y$-axis depicts the completeness of the data at different $G$ magnitudes, and the color scale indicated by the colorbar represents the corresponding $M_{10}$ (median magnitude of sources with {\tt astrometric\textunderscore matched\textunderscore transits} $\leq 10$) values. 
The blue vertical dashed line indicates $G = 18$\,mag, which is the maximum magnitude we investigate in our study. 
Based on the model results, we can consider {\it Gaia} data with $G < 18$\,mag within the cluster radius to be complete.}
\end{figure*}

We conduct the completeness check by selecting all {\it Gaia} sources within $r_\mathrm{cl}$ of each Snake cluster and applying the \citet{cantat2023empirical} model. We show the results in Figure \ref{fig:completeness_cluster}.
We observe that, according to the model results, data completeness for all stars within the $r_\mathrm{cl}$ range has reached completeness for $G<19$ mag except UBC\,7.
Considering that our study focuses only on stars with $G < 18$\,mag, we can reasonably conclude that the initial {\it Gaia} data within the scope of our research is complete. Therefore, we conclude that the samples within these regions are not significantly affected by selection biases originating from $\it{Gaia}$ itself.

\begin{figure}
\centering
\includegraphics[width=1.0\linewidth]{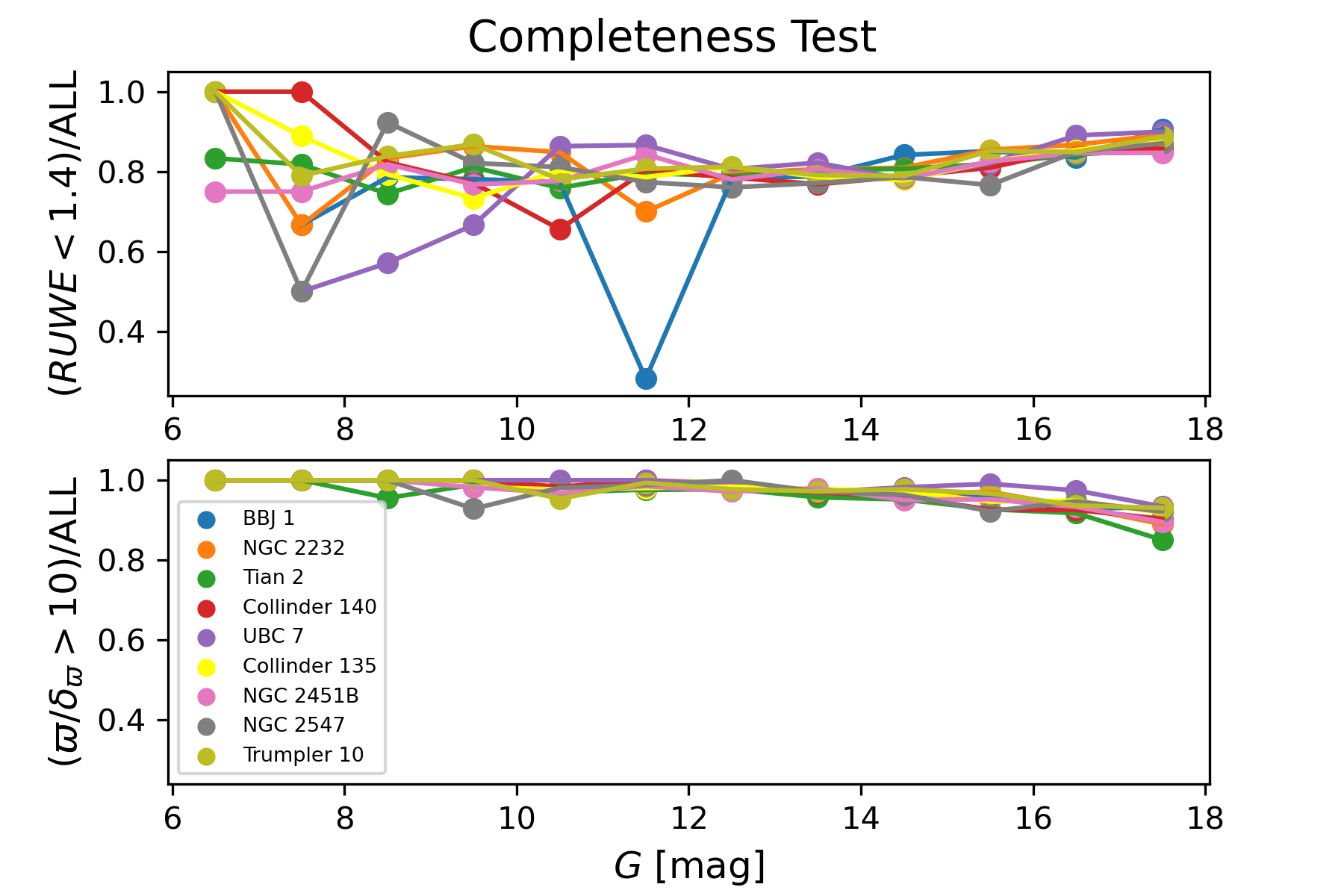}
\caption{\label{fig:completeness}Variation of data completeness along $G$ for all {\it Gaia} sources within the $r_\mathrm{cl}$ of the different Snake open clusters. The upper panel compares the effect of the RUWE restriction (RUWE $< 1.4$) on data completeness and the lower panel compares the effect of the {\tt parallax\_over\_error} restriction ($\varpi/\delta_{\varpi} > 10.0$) on data completeness. The $x$-axis represents different $G$ intervals. The $y$-axis in the upper panel represents the ratio of the number of all {\it Gaia} data sources with RUWE $< 1.4$ to the number of all {\it Gaia} data sources without the RUWE restriction. The $y$-axis in the lower panel represents the ratio of the number of all {\it Gaia} data sources with $\varpi/\delta_{\varpi} > 10.0$ to the number of all {\it Gaia} data sources without {\tt parallax\_over\_error} restrictions. Each open cluster is represented by a different colored data point. 
Limiting RUWE $< 1.4$, except for BBJ\,1, generally maintains completeness around 0.8 for $G > 9$\,mag data, while completeness is consistently at or above 0.9 for limiting $\varpi/\delta_{\varpi} > 10.0$.}
\end{figure}

After verifying the completeness of the {\it Gaia} data within the regions of our Stellar Snake clusters, we next move on to investigate the second aforementioned factor influencing the MF, this being due to the criteria we applied to select high quality Stellar Snake members.
We consider the selection criteria that we follow as presented in \cite{wang2022stellar} to assess their potential impact on data completeness.
For the following two tests, we explore the completeness changes 
by comparing all {\it Gaia} data within each $r_\mathrm{cl}$ with and without applying the two different selection criteria.
Specifically, to assess the impact of RUWE and {\tt parallax\_over\_error} on data completeness, we divide all {\it Gaia} sources within $r_\mathrm{cl}$ for each open cluster into two groups. The first group includes all {\it Gaia} data within the radius without any restrictions, while the second group consists of sources with restrictions only on RUWE $< 1.4$ and those with limitations only on $\varpi/\delta_{\varpi} > 10.0$. We compare the ratio of the number of sources in different $G$ magnitude intervals between the two groups to evaluate the effect of the RUWE and  {\tt parallax\_over\_error} restriction on data completeness. The results are shown in the Figure \ref{fig:completeness}.

Specifically, as shown in the lower panel of Figure \ref{fig:completeness}, we investigate our selection criteria on the relative precision criterion {\tt parallax\_over\_error} ($\varpi/\delta_{\varpi} > 10.0$). 
This condition aims to ensure more accurate parallax measurements, thereby obtaining more reliable member stars. 
Sources with higher {\tt parallax\_over\_error} generally correspond to more distant sources and fainter sources. Considering that the distance to the Snake is approximately 300 pc and $G$ less than 18\,mag, we expect that by applying the criterion of $\varpi/\delta_{\varpi} > 10.0$, we are able to improve the overall data quality without compromising the completeness of the sample, thus aiding in the identification of reliable member stars. 

We now test our expectation that the completeness remains uncompromised by this criterion. For all {\it Gaia} sources within $r_\mathrm{cl}$ of each cluster, we search for changes in the completeness under different $G$ magnitudes when restricting $\varpi/\delta_{\varpi} > 10.0$.
We observe, as shown in Figure \ref{fig:completeness}, that as $G$ decreases, the differences between the two data sets also decrease and eventually approach unity. This indicates that even under the worst-case scenario, the completeness remains above 0.8. Therefore, we conclude that $\varpi/\delta_{\varpi} > 10.0$  does not significantly impact our results. 

\begin{figure*}
\centering
\includegraphics[width=1.0\linewidth]{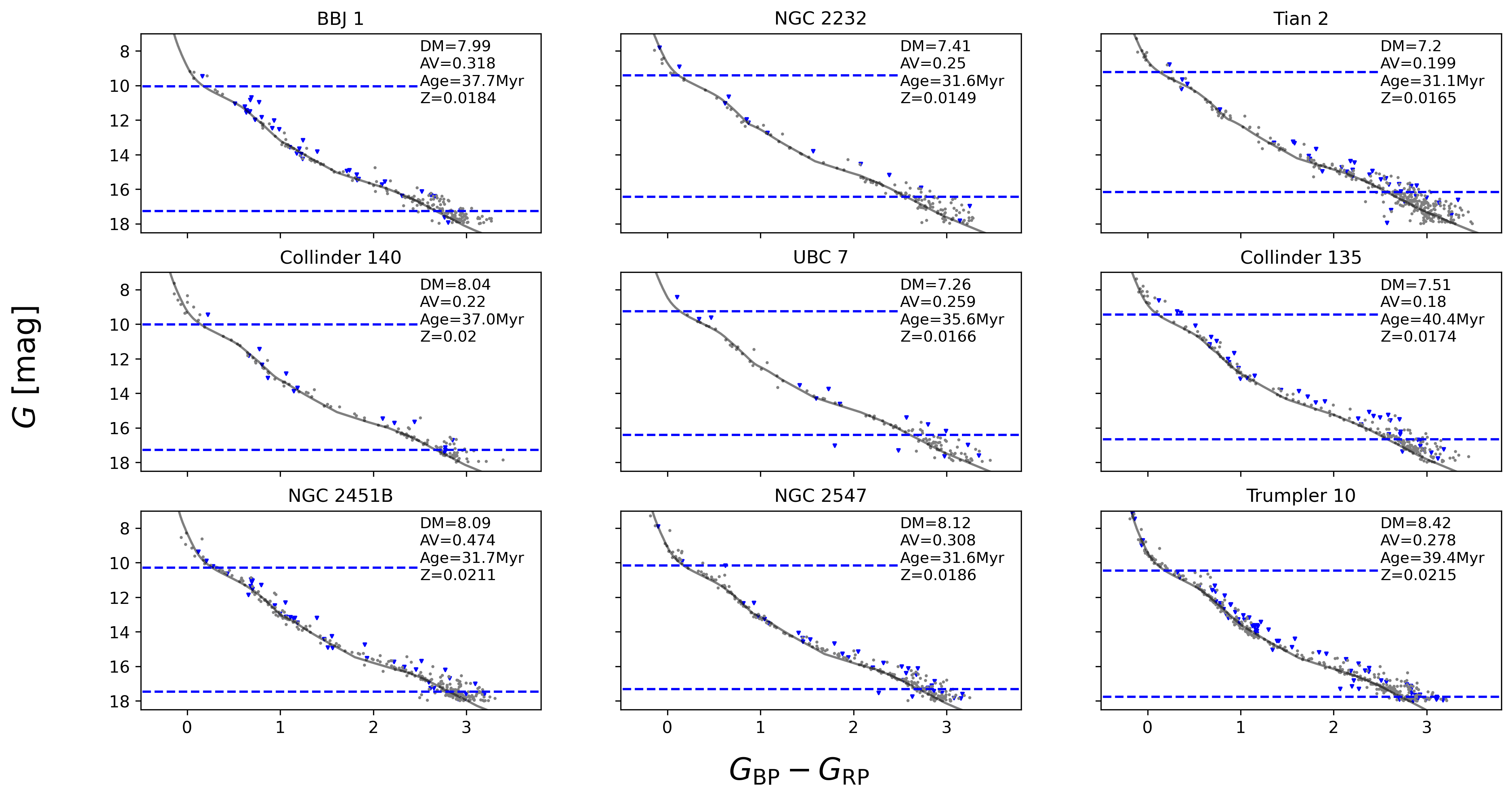}
\caption{\label{fig:CMD}CMDs for the 9 open clusters of the Stellar Snake's Parts I and II. The gray points represent the member stars selected when restricting RUWE $< 1.4$, the blue inverted triangles represent the additional member stars that are selected when we do not impose the RUWE $< 1.4$ restriction (Section \ref{subsection:final_members}). The gray curves represent the best-fitting isochrones. 
\textbf{The blue horizontal dashed lines demark the apparent magnitude range for single stars within the mass range $[0.5-2.0\, \mathrm{M}_\odot]$. }
The parameters related to the isochrone fitting are shown in the upper-right corner of each panel. 
DM, Age, $A_V$ and $Z$ represents the best-fitting distance modulus,  age,  extinction and metallicity.
These clusters have ages ranging from $30-40$ Myr and metallicities slightly higher than solar metallicity.
\label{CMD}}
\end{figure*}

Our Stellar Snake members are also limited to sources with RUWE $< 1.4$. We now consider this additional limiting condition and its impact on data completeness.
The RUWE is a metric defined in \cite{lindegren2018gaia} that quantifies the goodness-of-fit of the astrometric solution provided by {\it Gaia}. Sources with high RUWE values may be indicative of either of two scenarios: 1) the source is an unresolved binary system that {\it Gaia} did not resolve, or 2) the quality of the source's observation is poor and has higher overall errors.


In the upper panel of Figure \ref{fig:completeness}, we find that in most cases, the ratio between the two groups (with and without the RUWE restriction) remains above 0.7, indicating good data completeness. However, for the BBJ\,1 open cluster, we find an anomaly where the completeness ratio drops to around 0.3 at $G$ magnitudes around 11.5\,mag. This discrepancy might be attributed to a higher number of unresolved binary stars and poor observational quality in $G$ magnitudes around 11.5\,mag, resulting in significant data incompleteness. 

To obtain more reliable results, we further divide the member star data into two groups of preliminary samples using the {\tt ROCKSTAR} algorithm: one with RUWE < 1.4 (Sec. \ref{subsection:final_members}), and the other group 
with the same initial conditions (Sec. \ref{Snake members}) and $r_\mathrm{cl}$ and parallax selections (Sec. \ref{subsection:final_members}) 
but without the RUWE restriction. In the following sections, we analyze the data by dividing it into two groups to discuss the impact of RUWE restrictions on data completeness and the potential errors introduced when RUWE is not considered, which can lead to unreliable data.

Finally, we obtain data for two groups of member stars of the open clusters. The distribution of the two groups of member stars of the open clusters, along with the best-fitting isochrones obtained from {\tt ASteCA}, are displayed in the CMD as shown in Figure \ref{fig:CMD}. The blue inverted triangles represent the stars that are additionally selected when we do not impose the RUWE restriction compared to the case with the RUWE restriction (gray markers). For the BBJ\,1 open cluster, we observe a significant difference between the two groups around $G = 11.5$\,mag. Specifically, the RUWE restriction did not identify any member stars in this region, while the sample without RUWE restriction included 10 additional member stars. This discrepancy highlights the issue of data completeness, as the completeness ratio between the two groups is only 0.3 in this region. Despite the completeness discrepancy of BBJ\,1, we find that BBJ\,1 and all the other eight Stellar Snake clusters that we analyze have best-fitting ages and metallicities that are consistent with our expectations, such that the ages range from $30-40$ Myr and the metallicities are slightly higher than solar metallicity.

\section{Simulated Color-Magnitude Model}\label{sec:model}
One of the commonly used methods to determine the MF is by obtaining the mass of coeval stars through their positions in the color-magnitude diagram \citep[CMD,][]{sollima2019stellar,hallakoun2021bottom,pang2021disruption}. 
Taking into account the relatively small number of member stars in open clusters, directly calculating the mass of each star and fitting the corresponding histograms using polynomials to compute the MFs faces certain difficulties.
The process of binning data when fitting a distribution often introduces unnecessary ambiguity in selecting the bins, which can result in biases in the inferred distribution. This is especially true when the number of samples in each bin is unequal, as demonstrated by studies such as \citet{apellaniz2005numerical} and \citet{maschberger2009estimators}.

On the other hand, only as an approximation can we directly treat all cluster member stars as single stars and calculate their MFs.  More precisely, we need to consider the impact of binary stars, as binary proportions are generally higher in open clusters. If we treat multiple lower-mass stars in binary systems as single stars with higher mass, this results in a flattening of the MFs. Therefore, for accurate MF measurements, we need to consider the influence of binaries. Fortunately, for binary systems, due to their overall luminosity being brighter and redder compared to single stars, we can identify these binaries in the CMD, usually distributed at the red side of the main sequence \citep[e.g.,][]{romani1991limits,bolte1992ccd,rubenstein1997hubble,bellazzini2002deep,clark2004blue,richer2004hubble,zhao2005hubble,milone2009multiple}. Therefore, we can clearly distinguish these binary systems in the CMD, treating them separately as binary systems, thereby obtaining a more accurate MF. 
\textbf{This method has been applied in numerous cluster studies and has yielded good results 
\citep[e.g.,][]
{sollima2007fraction,milone2009multiple,milone2016binary,ji2015binary,li2020modeling,almeida2023revisiting}.}

Taking into account the aforementioned factors, we build a simulated CMD model using the MCMC algorithm to determine the MF, binary fraction ($f_b$), and binary mass ratio distribution ($\gamma$) within each cluster. This approach helps avoid biases introduced by binning the data and also considers the impact of binaries on the MFs. 


We now proceed to model the color-magnitude diagram (CMD) based on prior knowledge such as the initial mass function (IMF), binary fraction, and mass-ratio distribution. 
This CMD model, synthesized from theoretical recipes (Section\,\ref{Sim star}), will then be meticulously compared with the observed CMD to determine the model parameters (Section\,\ref{bayesian}).

\subsection{Synthetic stars}\label{Sim star}
In order to study the properties of our open clusters, we generate synthetic stars that resemble the observed stellar populations following the proceeding steps.
\begin{enumerate}
\item We generate a population of synthetic single stars and binary systems within a specific mass range by utilizing the mass function formula,
\begin{equation}
\xi_{\mathrm{MF}} = \frac{dN}{dM}
      \propto M^\alpha,
\end{equation}
where $M$ represents the stellar mass and $dM/dN$ is the number of stars within the mass range of $M$ to $M+dM$. The MF power-law index is represented as $\alpha$. 

We generate $100,000 \times (1 - f_\mathrm{b})$ synthetic single stars with masses $M_i$ and $100,000 \times f_\mathrm{b}$ binary systems with primary stellar masses $M^\mathrm{p}_i$ within the specified mass range.
\item 
We obtain the best-fit PARSEC isochrones for the star clusters by using the optimal metallicity and age values provided in Table \ref{tab:Parameters}.
These isochrones provide the corresponding range of colors and magnitudes in the CMD for stars within the specific mass range of the cluster.


\item Using the mass $M_i$ of each synthetic star, we perform interpolation to assign appropriate intrinsic colors ${\rm Col}_{0,i}$ and absolute magnitudes $M_{G_0,i}$ based on the best-fitting isochrone. Similarly, for the binary systems, we assign intrinsic colors ${\rm Col}^{p}_{0,i}$ and absolute magnitudes $M^{p}_{G_0, i}$ to the primary stars using their masses $M^{p}_i$.
\item For the single stars, we add the corresponding distance modulus DM and extinction $A_G$ to each single star, as well as the extinction in the $G_{\rm BP\_RP}$ color index $A_{\rm BP\_RP}$. Here, $A_G$ is calculated as $A_V\times 0.836$, and $A_{\rm BP\_RP}$ is calculated as $0.449 \times A_V$ as provided by \citet{cardelli1989relationship}. 
These extinction coefficients are consistent with those used in {\tt ASteCA}.
By incorporating the DM, $A_G$, and $A_{\rm bp\_rp}$ values, we obtain the mock observed colors ${\rm Col}_{1,i}$ and apparent magnitudes $G_{1,i}$ for each single synthetic star, accounting for the effects of distance and extinction.

\item  For the errors in ${\rm Col}_{1,i}$ and $G_{1,i}$, we divide them into multiple small bins based on the errors on the real data,  $e_{\rm Col}$ (color error) and $e_{G}$ (magnitude error). We use the median value of the error within each bin as the representative error for the synthetic stars within that bin. 
Considering the systematic errors introduced during the processing of {\it Gaia} data, uneven distribution of extinction among cluster member stars, various physical factors inherent to stars, etc., the actual distribution of member stars may be more scattered than the simulated star distribution obtained through observational errors. To simulate more realistic observational results, we follow the method proposed by \citet{li2022mimo} and add an additional scatter of 0.01\,mag to the errors to mimic the observations.
Then we add random errors with standard deviations  $e_{\rm Col}$ and $e_{G}$ to each synthetic star within the corresponding bins, to obtain values ${\rm Col}^*_i$ and $G^*_i$, representing the final mock observed colors and magnitudes of the synthetic single stars.
\item For binary stars, we only consider binary systems with mass ratios ($q = M1/M2$) greater than 0.3. This is because unresolved binary systems with $q < 0.3$ are difficult to distinguish from single stars. To assign masses to the secondary stars ($M^{s}_i$) for each primary star, we use a given value of $\gamma$, which is the power-law slope in the equation,
\begin{equation}
\xi_{q} = \frac{dq}{dN}
      \propto q^\gamma,
\end{equation}
which represents the distribution of binary mass ratios. We generate the masses of the secondary stars based on this distribution. Subsequently, we perform interpolation using the mass of the secondary star $M^{s}_i$ to determine the corresponding color ${\rm Col}^{s}_{0,i}$ and absolute magnitudes $M^{s}_{G_0,i}$ of the secondary star for the binary systems.
\item For both the primary and secondary stars in the binary systems, we synthesize their absolute magnitudes $M^{\mathrm{tot}}_{G_0, i}$ and colors Col$_\mathrm{tot}$ by using the formula,
\begin{equation}
M^{\mathrm{tot}}_{G_0, i} = -2.5
\log{(10^{(-0.4\times M^{p}_{G_0, i})}+10^{(-0.4\times M^{s}_{G_0, i})})}.
\end{equation}

After synthesizing the magnitudes and colors, we add the corresponding errors, DM, and $A_V$ to obtain the synthetic magnitudes $G^*_{\mathrm{tot},i}$ and colors ${\rm Col}^*_{\mathrm{tot},i}$ of the binary stars. The errors, DM, and $A_V$ account for uncertainties in the observations and the effects of distance and extinction.
\end{enumerate}

\begin{figure}
\centering
\includegraphics[width=1.0\linewidth]{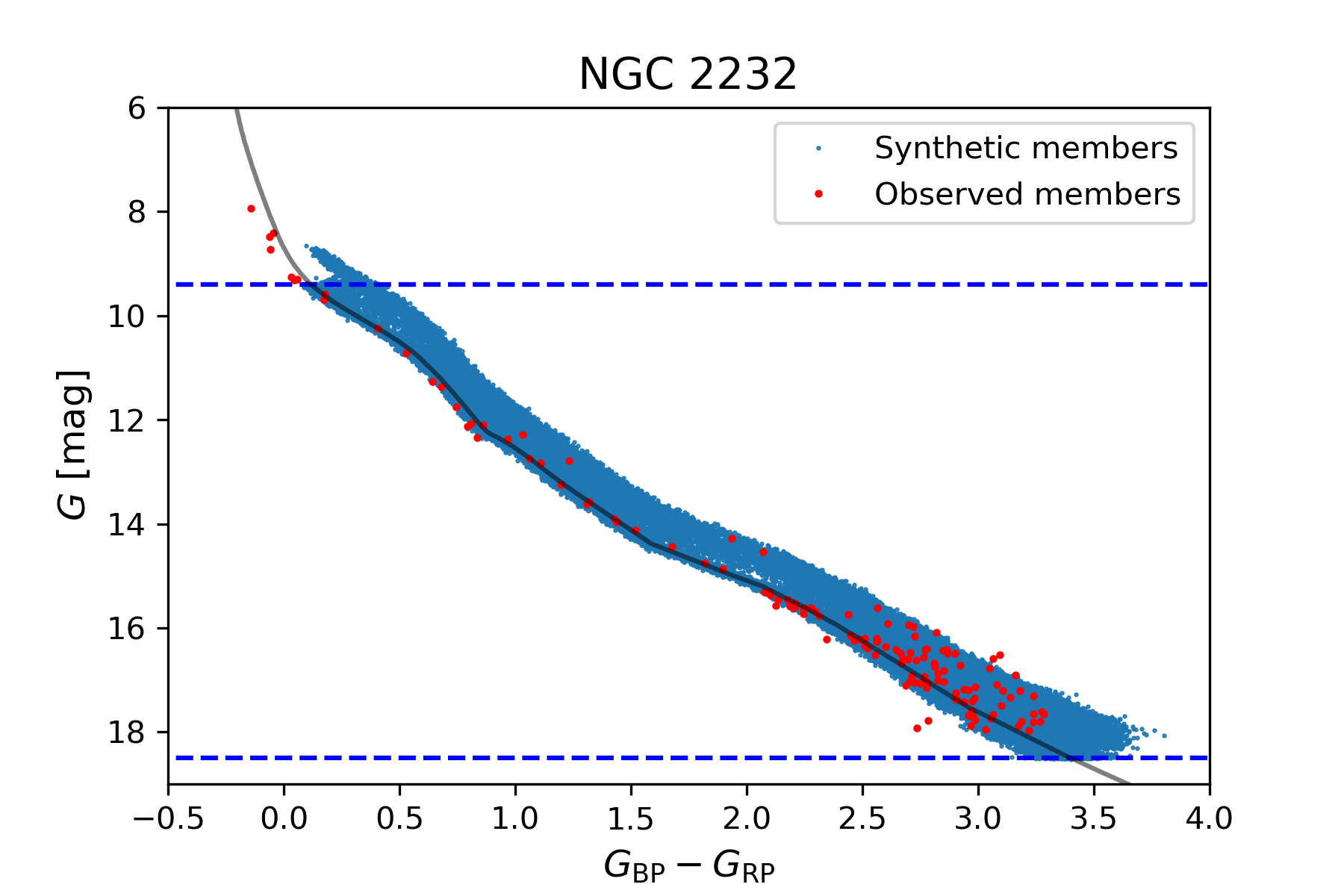}
\caption{\label{fig:Mock compare}Comparison between the synthetic and observed member stars for the Stellar Snake open cluster NGC\,2232. The gray curve represents the best-fitting isochrone. The blue dots represent the 100,000 synthetic stars generated with specific $\alpha$, $f_b$, and $\gamma$ within a given mass range [$M_1$, $M_2$] (where $M_2$ > $M_1$). The blue horizontal dashed lines demark the apparent magnitude range for single stars within the mass range [$M_1$, $M_2$]. The red dots represent the {\it Gaia} DR3 observed member stars of NGC\,2232.}
\end{figure}

With these steps, we obtain the colors (${\rm Col}^*_i$, ${\rm Col}^*_{\mathrm{tot},i}$) and magnitudes ($G^*_i$, $G^*_{\mathrm{tot},i}$) for the synthetic single stars and binary stars. These values are used to characterize the synthetic stellar samples generated based on the corresponding $\alpha$, $f_b$, and $\gamma$ parameters. Figure \ref{fig:Mock compare} shows a comparison between the synthetic stars and the observed Stellar Snake open cluster member stars for NGC\,2232.

\subsection{Bayesian Statistical Inference}\label{bayesian}

We treat $p_i({\rm Col}_i,G_i | \Theta)$ as the probability density distribution for each member star, where $\Theta$ represents the set of parameter describing the distribution of the star cluster, which includes $\alpha$, $f_b$, and $\gamma$. The prior probability distributions for these parameters are listed in Table \ref{tab:prior}.

\begin{table}
\centering
\caption{\label{tab:prior} Bayesian priors used to generate synthetic star clusters.}
\scalebox{0.85}{
\begin{tabular}{c|cc}
\hline
 & Range & Description \\
\hline
$\alpha$ & $U \sim [-5, 0]$ & Power-law index of mass function \\
$f_b$ & $U \sim [0, 1]$ & Fraction of binaries with $q > 0.3$ \\
$\gamma$ & $U \sim [-10, 10]$ & Power-law index of the binary mass ratio distribution \\
\hline
\end{tabular}}
\end{table}

To obtain the probability density distribution, we use the kernel density estimation (KDE) method to analyze the colors (${\rm Col}^*_i$, ${\rm Col}^*_{\mathrm{tot},i}$) and magnitudes ($G^*_i$, $G^*_{\mathrm{tot},i}$) of 100,000 synthetic stars generated within the specific mass intervals in the CMD. For the selection of bandwidth and kernel function, we perform a 10-fold cross-validation using the {\tt GridSearchCV} module from {\tt scikit-learn}\footnote{\label{link4} \href{https://scikit-learn.org/stable/}{https://scikit-learn.org/stable/}}. We consider different bandwidth and kernel function combinations, calculating the scores for each combination to determine the optimal bandwidth and kernel function.

Subsequently, we apply the true data to the KDE model to obtain the probability density $p_i$ for each star. The final likelihood is then computed as follows:

\begin{equation}
\mathcal{L}({\rm Col}_i,G_i|\theta)=\prod_i p_i({\rm Col}_i,G_i|\theta).
\end{equation}

According to Bayes' theorem, we obtain the posterior probability density distribution of $\Theta$ given the observed data (${\rm Col}_i, G_i$):
\begin{equation}
P(\Theta\mid {\rm Col}_i,G_i)\propto \mathcal{L}({\rm Col}_i,G_i\mid\theta)\times\Pi(\theta),
\end{equation}
where $P(\Theta\mid {\rm Col}_i,G_i)$ is the posterior probability density distribution of the parameter set $\Theta$, $\mathcal{L}({\rm Col}_i,G_i\mid\theta)$ is the likelihood function representing the probability density distribution of the observed data given the parameter set $\Theta$, and $\Pi(\theta)$ is the prior probability density distribution of the parameter set $\Theta$.

 We utilize the MCMC algorithm within the {\tt emcee}\footnote{\label{link3} \href{https://emcee.readthedocs.io/en/stable/index.html}{https://emcee.readthedocs.io/en/stable/index.html}} \citep{foreman2013emcee} package, employing 120 walkers and 3000 steps to sample the fitting parameters and obtain the posterior probability density function of the parameters. Finally, we use the median (50th percentile) as the best-fit result and the 16th and 84th percentiles as the 1$\sigma$ uncertainty interval.

\section{RESULTS}\label{sec:result}
In order to make use of the model built in the previous section, we utilize the fundamental parameters from Table \ref{tab:Parameters} for each individual open cluster calculated by {\tt ASteCA}. We download the appropriate PARSEC isochrones corresponding to the specified age and metallicity values for each cluster. By inputting the calculated extinction and distance modulus, we use our comprehensive model described in Section\,\ref{sec:model} to determine the masses corresponding to the brightest and faintest stars within each cluster. Subsequently, we calculate the respective mass functions and binary ratios for the Stellar Snake star clusters and present the intriguing trends in Section\,\ref{subsec: result}, and provide explanation for these results in Section\,\ref{subsec:explain}.

\subsection{Variations in the Mass Function}\label{subsec: result}

In order to facilitate meaningful comparisons between different open clusters within the Stellar Snake, \textbf{we ensure that the fitting range for each star cluster falls within the same mass interval. Taking into account that at $G = 18\,$mag, the minimum mass for these clusters is approximately $0.45\, \mathrm{M}_\odot$ (Trumpler\,10), and the lower limit for the maximum mass for all clusters is around $2.0\, \mathrm{M}_\odot$ (UBC\,7).}
We thus calculate the mass functions for each cluster in the same mass range of $[0.5-2.0\, \mathrm{M}_\odot]$, separately for the two datasets, one with RUWE restricted and the other without the RUWE restriction. By selecting this mass range, we eliminate members of the clusters; but this approach using the same mass range eliminates the dim cluster members with higher uncertainties and allows us to better compare the mass function relationships of different open clusters within the Snake region. The final results are presented in Table \ref{tab:result}.

\begin{table*}
\caption{\label{tab:result}Fitting results from our simulated CMD model for each Stellar Snake open cluster.}
\begin{tabular}{c|c|c|c|c|c|c|c|c}
Cluster      & $G$ range\, [mag] & Mass range & $\alpha$     & $f_b$   & $\gamma$  & ${\alpha}^*$ & ${f_b}^*$ & ${\gamma}^*$ \\\hline
BBJ\,1  & 10.04$-$17.27        & 0.5$-$2.0\,$\mathrm{M}_\odot$    & $-3.57^{+0.36}_{-0.37}$ & $0.34^{+0.06}_{-0.05}$ & $0.20^{+0.79}_{-0.78}$ &  $-2.82^{+0.28}_{-0.33}$  &  $0.39^{+0.06}_{-0.07}$   &  $0.76^{+1.02}_{-0.82}$  \\
NGC\,2232  & 9.40$-$16.43     & 0.5$-$2.0\,$\mathrm{M}_\odot$    & $-3.01^{+0.34}_{-0.40}$ & $0.23^{+0.08}_{-0.06}$ & $0.78^{+1.18}_{-0.96}$ &  $-2.94^{+0.36}_{-0.34}$    &  $0.24^{+0.06}_{-0.05}$   &  $1.80^{+1.13}_{-1.25}$  \\
Tian\,2    & 9.23$-$16.15   & 0.5$-$2.0\,$\mathrm{M}_\odot$    & $-2.85^{+0.36}_{-0.33}$ & $0.45^{+0.09}_{-0.07}$ & $-1.06^{+0.96}_{-1.07}$ &    $-2.88^{+0.35}_{-0.41}$ &  $0.51^{+0.10}_{-0.08}$    &   $-0.69^{+0.72}_{-0.74}$ \\
Collinder\,140 & 10.02$-$17.27  & 0.5$-$2.0\,$\mathrm{M}_\odot$    & $-2.70^{+0.32}_{-0.42}$ & $0.38^{+0.07}_{-0.07}$ & $-1.16^{+0.85}_{-0.81}$ &  $-2.72^{+0.32}_{-0.39}$  &   $0.43^{+0.08}_{-0.09}$  &  $0.01^{+1.16}_{-1.05}$  \\
UBC\,7   & 9.26$-$16.40       & 0.5$-$2.0\,$\mathrm{M}_\odot$    & $-2.27^{+0.46}_{-0.41}$ & $0.36^{+0.10}_{-0.09}$ & $-2.89^{+1.03}_{-1.04}$ &  $-2.27^{+0.41}_{-0.44}$   &  $0.43^{+0.08}_{-0.09}$   &  $-1.45^{+1.03}_{-0.78}$  \\
Collinder\,135 & 9.44$-$16.67  & 0.5$-$2.0\,$\mathrm{M}_\odot$    & $-2.15^{+0.25}_{-0.28}$ & $0.54^{+0.07}_{-0.07}$ & $-2.34^{+0.46}_{-0.56}$ &   $-2.25^{+0.37}_{-0.37}$  &    $0.50^{+0.09}_{-0.06}$  &   $-1.04^{+0.65}_{-0.69}$ \\
NGC\,2451B   & 10.29$-$17.47   & 0.5$-$2.0\,$\mathrm{M}_\odot$    & $-2.54^{+0.34}_{-0.22}$ & $0.42^{+0.06}_{-0.08}$ & $-1.04^{+0.92}_{-0.87}$ &  $-2.45^{+0.37}_{-0.30}$  &  $0.46^{+0.07}_{-0.07}$   &  $-0.38^{+0.69}_{-0.72}$  \\
NGC\,2547   & 10.17$-$17.31    & 0.5$-$2.0\,$\mathrm{M}_\odot$    & $-2.37^{+0.25}_{-0.22}$ & $0.50^{+0.06}_{-0.06}$ & $-1.88^{+0.41}_{-0.48}$ &   $-2.40^{+0.25}_{-0.22}$  &  $0.49^{+0.06}_{-0.05}$   &  $-1.04^{+0.42}_{-0.43}$ \\
Trumpler\,10 & 10.45$-$17.75   & 0.5$-$2.0\,$\mathrm{M}_\odot$    & $-2.33^{+0.26}_{-0.19}$ & $0.34^{+0.04}_{-0.05}$ & $-0.46^{+0.64}_{-0.62}$ &   $-2.30^{+0.22}_{-0.29}$  &   $0.42^{+0.05}_{-0.05}$   &  $0.17^{+0.48}_{-0.48}$ 
\\\hline
\end{tabular}
\begin{minipage}{\textwidth}
\raggedright
\textbf{Note: }The values of $\alpha$, $f_b$, and $\gamma$ represent the mass function, binary fraction ($q>0.3$), and binary mass ratio distribution, respectively, obtained for data restricted to RUWE $< 1.4$ within the mass range of 0.5 to 2.0\,M$_\odot$. On the other hand, ${\alpha}^*$, ${f_b}^*$, and ${\gamma}^*$ correspond to the same except for the data when RUWE is unrestricted.
\end{minipage}
\end{table*}

From the results obtained within the mass range of 0.5 $\mathrm{M}_\odot$ to 2.0 $\mathrm{M}_\odot$, we observe that the binary fraction ($q > 0.3$) for most of the clusters is concentrated between 30\% to 50\%, and $\gamma$ is predominantly negative, indicating that the majority of members in these open clusters have lower-mass companions in binary systems. Regarding their mass functions, the majority of clusters follow a mass function close to the Kroupa initial mass function with a value of $\alpha\sim -2.3$. However, there are also four clusters (Collinder\,140, Tian\,2, NGC\,2232, BBJ\,1) with a value of $\alpha<-2.3$, indicating that these clusters have a higher proportion of low-mass stars and relatively fewer high-mass stars.

\begin{figure*}
\centering
\includegraphics[width=1.0\linewidth]{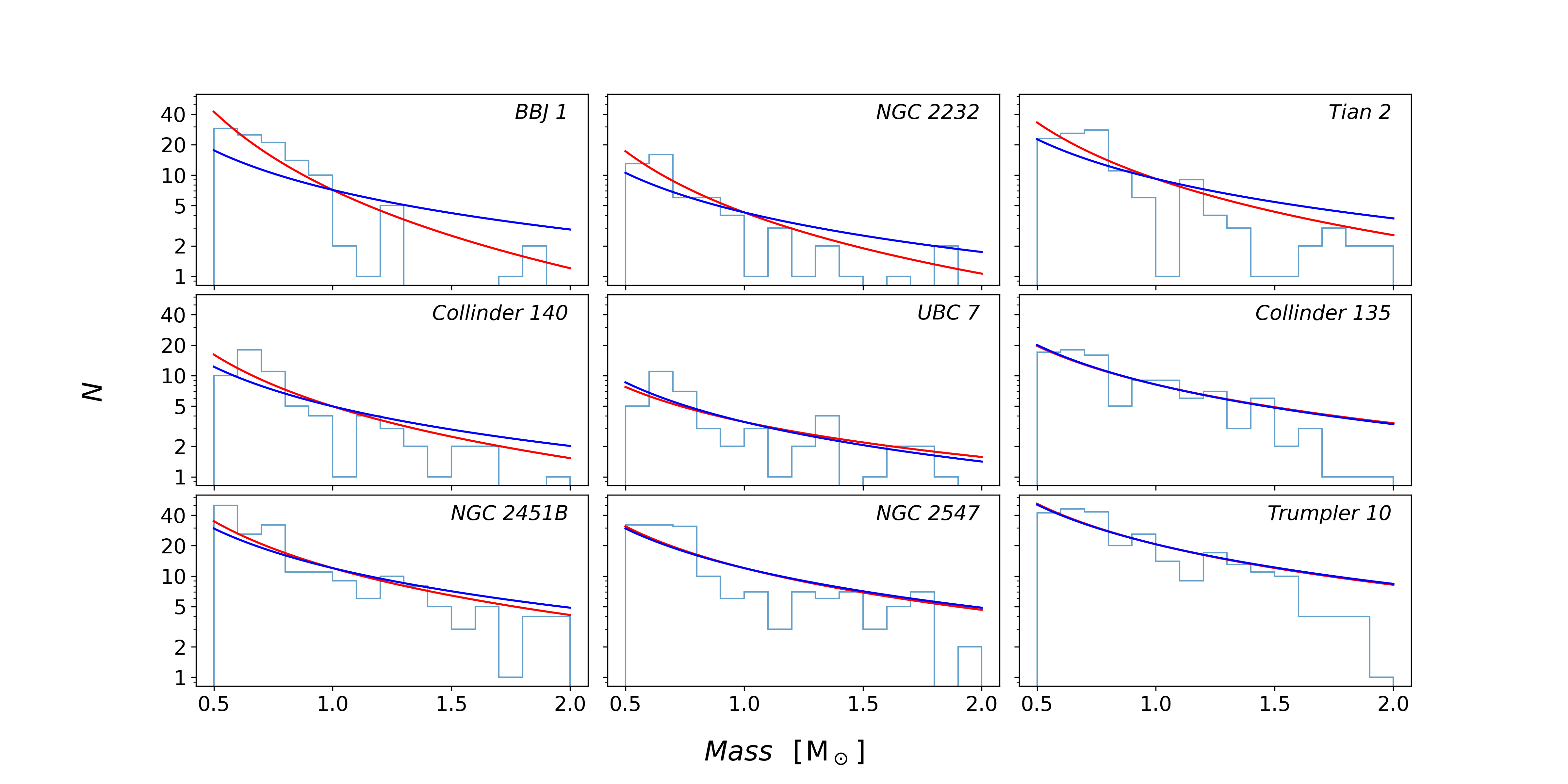}
\caption{\label{fig:Mass_N}Verification of results for Snake clusters with RUWE $<1.4$ in the mass range of 0.5\,$\mathrm{M}_\odot$ to 2.0\,$\mathrm{M}_\odot$. The $x$-axis represents the number of stars ($N$), and the $y$-axis represents the mass. The histogram presents the distribution of member stars within various mass intervals calculated directly from the CMD and the best-fit PARSEC isochrone found by {\tt ASteCA}. The red line corresponds to the best-fit result from our model  while the blue line represents the Kroupa IMF value for masses greater than 0.5\,$\mathrm{M}_\odot$ ($\alpha = -2.3$).}
\end{figure*}

To further validate the reliability of our results, we employ {\tt XGBoost} \citep{chen2016xgboost} to compare the member stars' $G_{\rm BP}-G_{\rm RP}$ and $G$ with the corresponding PARSEC isochrones, taking into account the extinction and
distance modulus, to directly compute the mass of the member stars. Subsequently, we calculate the histogram for the computed masses within the specified mass range and compare it to the best-fit slope obtained from our simulated CMD model (Figure \ref{fig:Mass_N}). 
Specifically, for BBJ\,1, NGC\,2232, Collinder\,140, and Tian\,2, where the results deviate from the canonical Kroupa IMF, as shown in Figure \ref{fig:Mass_N}, we clearly observe that the fitting results (red curve) better match the histogram distribution compared to the Kroupa IMF (blue curve). In contrast, the other clusters exhibit results similar to the Kroupa IMF. 
Thus, we validate the reliability of our simulated CMD model and the fitting results by comparing them with the histograms.


Comparing the results show in Table \ref{tab:result} for the two sample groups, we can see that, except for BBJ\,1, the results of the other eight open clusters are very similar. Taking into account the corresponding errors,  the results of the two groups are consistent. However, for BBJ\,1, as previously indicated in the upper panel of Figure \ref{fig:completeness}, the impact of RUWE on its completeness is significant, leading to substantial differences in its results. The difference in $\alpha$ obtained from the two sample groups is also beyond their error ranges, indicating two distinct results. Therefore, we can conclude that whether RUWE is restricted has a significant impact on the results for BBJ\,1, while for the other clusters, it only leads to minor differences.

\begin{figure*}
\centering
\includegraphics[width=1.0\linewidth]{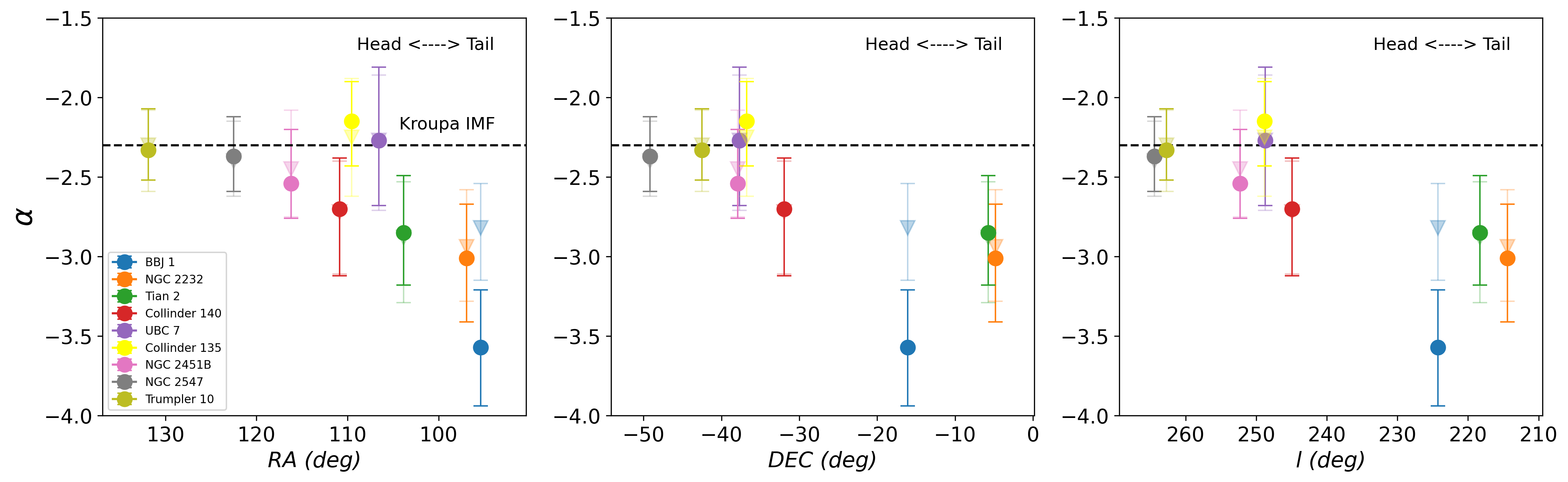}
\caption{\label{fig:result}Variation of the MF slopes $\alpha$ for member stars of each cluster in the range of 0.5 to 2.0 solar masses, along coordinates RA, DEC and $l$. Each open cluster is represented with a different color (as indicated in the legend). Circles represent the results obtained from our data sample with RUWE $<1.4$, and inverted triangles represent the results from the control group where we do not apply the RUWE restrictions. The horizontal dashed line represents the Kroupa initial mass function slope ($\alpha = -2.3$) for masses greater than 0.5 solar masses. Except for BBJ\,1, the results agree between our data sample and the control group. Intriguingly, we find a trend in $\alpha$ along
RA, DEC, and $l$ from the head to the tail of the Stellar Snake (the head/tail direction is indicated in the upper right of each panel).}
\end{figure*}

Intriguingly, we note a decreasing trend in the parameter $\alpha$ from the head-to-tail of the Snake. The coordinates RA, DEC, and $l$ effectively trace the length of the Snake from head-to-tail, therefore, the parameter $\alpha$ exhibits a significant correlation with RA, DEC, and $l$ as show in Figure \ref{fig:result}. 
Specifically, for the circle markers (RUWE restrictions), as RA decreases, $\alpha$ becomes more negative. However, it is worth noting that BBJ\,1 does not entirely follow this pattern in the direction of DEC and $l$. Except for BBJ\,1, there seems to be a trend in DEC and $l$, where $\alpha$ becomes more negative as DEC increases and $l$ decreases. 
As for the triangle markers (without RUWE restrictions), 
apart from BBJ\,1, the triangle markers are similar to the circle markers. Due to the variability in $\alpha$ for BBJ\,1, these triangle markers exhibit more pronounced trends in DEC and $l$, where $\alpha$ becomes more negative as the coordinate values increase and decrease, respectively.
However, in RA, this trend is more pronounced only when BBJ\,1 is excluded. 
The distribution of the Snake's length along coordinate $b$ exhibits a U-shaped pattern, indicating that $b$ does not effectively capture the head-to-tail distribution of the Snake; therefore, no clear trend in $\alpha$ is evident along $b$.

\subsection{Proposed Explanations for the Results}\label{subsec:explain}

In Section \ref{subsec: result}, we calculate the results for two sets of samples: one with the restriction $\mathrm{RUWE} < 1.4$ and the other without this restriction. We find that the results for BBJ\,1 show significant differences between these two samples, consistent with what we find in Section \ref{completeness}, where BBJ\,1 exhibits a pronounced completeness issue at the brighter end. The results for the other clusters remain largely consistent between these two sets of samples. The noticeable completeness issue for BBJ\,1 when limiting RUWE $< 1.4$, particularly around $G = 11.5$\,mag, may be attributed to the overall $\it{Gaia}$ data quality in the observed region or potentially the presence of a higher number of unresolved binary stars in the vicinity of $G = 11.5$\,mag.

We also find that whether RUWE is restricted or not, the MFs of the member clusters in the Stellar Snake appear to exhibit variations along the length of the Snake. Additionally, the Kroupa IMF ($\alpha = -2.35$ for mass $ > 0.5 \,\mathrm{M}_\odot$) lies outside the error range for four of the Snake open clusters that have smaller $\alpha$ values, suggesting they might contain more low-mass stars within the mass range $[0.5-2.0\, \mathrm{M}_\odot]$.

A plausible explanation for this is that the clusters near the ``tail'' are relatively younger compared to those near the ``head'' \citep{wang2022stellar}. Additionally, the delayed formation of massive stars \citep{krause2020physics} leads to a lack of high-mass stars in the MF within the mass range $[0.5-2.0\, \mathrm{M}_\odot]$.
As star-forming regions evolve, they form more massive stars, thus the most massive stars tend to be younger, because they only form when the star formation rate 
of the whole region has increased sufficiently to sample the massive regions of the IMF, according to the simulations of \citet[][hereafter, VS17]{vazquez2017hierarchical}.
The formation of massive stars necessitates environments with higher gas density, and reaching such conditions takes time. Hence, this leads to a delayed formation of massive stars. 
This is exemplified in figure 7 of VS17, which shows the distribution of the cumulative masses of early-stage cluster stars over time as obtained through simulations. They show that the formation of massive stars commences only when the merger of group\,1 and group\,2 generates a new cluster with a higher star formation rate, which they denote as group\,12.
Therefore, massive stars form only when the star formation rate is sufficiently high, predominantly in high-density regions, and thus massive stars are concentrated near the center of the cluster. 

\begin{figure*}
\centering
\includegraphics[width=1.0\linewidth]{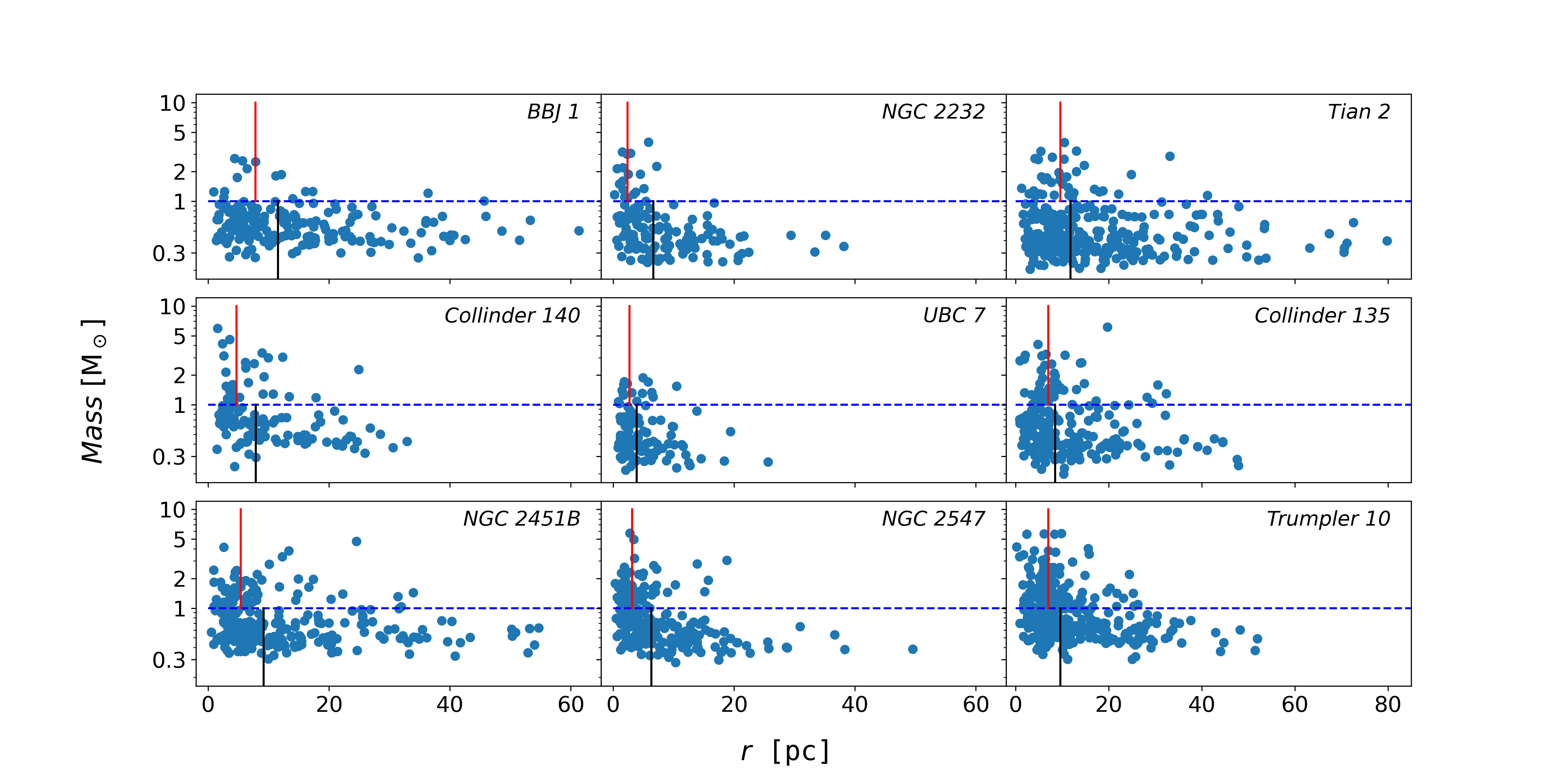}
\caption{\label{fig:R-Mass}Mass distribution of stars in the nine open clusters of the Stellar Snake. The distance $r$ of member stars from the cluster center is displayed on the $x$-axis and the mass of the member stars on the $y$-axis. 
\textbf{The horizontal blue dashed line represents the boundary at a mass of 1\,$\mathrm{M}_\odot$. The red and black vertical lines indicate the median distances from the cluster center $r$ for all member stars with mass > 1\,$\mathrm{M}_\odot$ and mass < 1\,$\mathrm{M}_\odot$. }}
\end{figure*}

In Figure \ref{fig:R-Mass}, we compare the masses of the Stellar Snake member stars in the nine open clusters and their distances from the cluster centers. The results closely resemble the model of VS17, with massive stars predominantly clustered near the center of the clusters, while low-mass stars exhibit a more dispersed distribution.
Some young clusters also appear "mass segregated," where the more-massive stars are more concentrated in the centers of clusters \citep{bonnell1998mass,de2002mass,gouliermis2004mass,chen2007mass}.
Considering that the Snake clusters have relatively young ages and are less likely to be influenced by significant dynamical evolution, we expect that the concentration of massive stars toward the center of the clusters is likely to be primordial.




This delayed formation of massive stars also implies that the age range of the massive stars is slightly smaller (and they are younger) than that of the low-mass stars, which begin to form since the onset of the star formation activity in the region. Equivalently, the mass range of the younger stars is larger and extends to higher masses than that of the older stars \citep{krause2020physics}.
Therefore, the delayed formation of massive stars suggests that, compared to the expected number produced by the overall stellar population, there is a scarcity of massive stars in the early phases of evolution in the ensembles of cores. This finding aligns with observations reported by \cite{povich2016rapid}.

\begin{figure}
\centering
\includegraphics[width=1.0\linewidth]{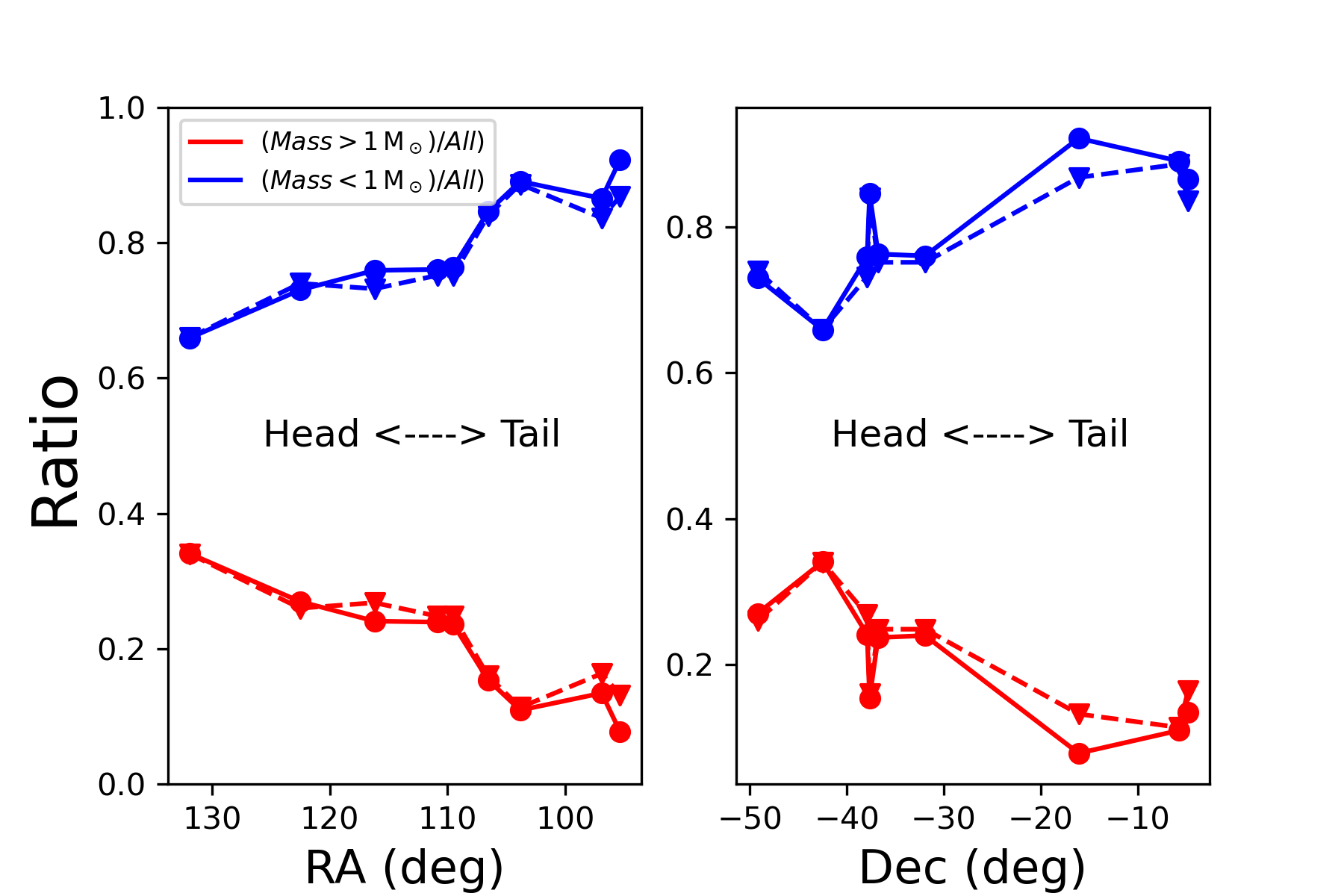}
\caption{\label{fig:Mass ratio}\textbf{Proportion of massive and low-mass stars within each star cluster as a function of the spatial position (RA, DEC) of each cluster along the Stellar Snake. The blue dashed line and circles represent the trend of the proportion of cluster member stars with masses less than 1\,$\mathrm{M}_\odot$ relative to all member stars, while the red dashed line and circles depict the trend of the proportion of cluster member stars with masses greater than 1\,$\mathrm{M}_\odot$ relative to all member stars. The dotted line and inverted triangles indicate the variation trend for the control group without applying RUWE restrictions. }}
\end{figure}

As mentioned by \cite{wang2022stellar}, the Snake is a hierarchically primordial structure and probably formed from a filamentary giant molecular cloud. Gravitationally bound stellar clusters in this structure arise naturally at localized high-density regions, while unbound associations form in situ at low density regions.
Therefore, the most plausible scenario for the phenomenon observed in Figure \ref{fig:result} is as follows, the older open clusters in the head of the Snake (e.g., Trumpler\,10, the physical pair of Collinder\,135 and UBC\,7, NGC\,2547, and NGC\,2451B) were firstly born in the giant molecular cloud around $30-40$ Myr ago, then followed the formation of the filamentary substructures in the tail of the Snake including four open clusters (e.g., Collinder\,140, BBJ\,1, and the physical pair of Tian\,2 and NGC\,2232) only a few Myrs later. 
However, due to the delayed formation of massive stars, when the clusters in the head of the Snake formed massive stars, the feedback in the form of stellar winds from the massive stars expelled the surrounding gas. This resulted in the filaments around the clusters in the head having a gas density insufficient to support the formation of massive stars, and star formation gradually ceased. Meanwhile, as the clusters in the tail began forming later, their formation was not sufficiently supported by gas due to the influence of stellar winds. This led to the lack of massive stars in the mass range of these relatively young clusters in the tail of the Snake.
\textbf{Figure \ref{fig:Mass ratio} provides evidence that the proportion of massive stars within a cluster decreases as a function of the position of the cluster along the head to the tail of the Snake.}

The Stellar Snake region exhibits a layered formation along the extent of its filamentary structure, with a significant spread in the properties of its open clusters along RA, DEC, and $l$. Interestingly, the ``head" of the Stellar Snake, corresponding to larger values of RA and $l$ (smaller DEC), appears to house relatively older open clusters. Conversely, the ``tail" of the Stellar Snake, characterized by smaller values of RA and $l$ (larger DEC), coincides with the four open clusters that do not conform to the Kroupa IMF. This observation seems plausible after taking into account 
the delay in the formation of high-mass stars, these relatively young clusters lack high-mass stars due to stellar feedback. 
This provides a reasonable explanation for the anomalous $\alpha$ observed in these open clusters.

The best-fit ages obtained from {\tt ASteCA} provide a rough estimate of the cluster's age; the error range is approximately $5$\,Myr. Such errors give an explanation to the lack of strong correlation between $\alpha$ and the best-fit age. The correlation between $\alpha$ and the spatial distribution of the clusters shows a much stronger trend.

\section{Discussion}\label{sec:discussion}

\subsection{Determination of Open Cluster Member Stars}
The selection of cluster members has always been a challenging aspect in stellar cluster studies \citep[e.g.,][]{lada2003embedded,sung2004initial,bouy2013dynamical,kuhn2014spatial}. Different clustering methods can yield different results \citep[e.g.,][]{hunt2021improving}, and the optimal method may vary depending on the research objectives. Determining the MF is particularly challenging, as obtaining the most precise MF results requires not only ensuring a large sample of cluster members but also avoiding the influence of field stars that might be introduced by relaxing selection criteria.

The comparison of different star clusters requires that the initial sample selection criteria and restrictions applied are as similar as possible for each cluster being compared. This ensures that the results obtained are most reliable. 

Taking these factors into consideration, we propose that the Stellar Snake, with its approximately uniform metallicity and relatively young age, provides an excellent environment for studying the MF. We use a combination of the {\tt ROCKSTAR} FoF method and {\tt ASteCA} to obtain the corresponding cluster members. Using FoF to filter the overall structure of the Stellar Snake helps ensure that cluster members are selected under uniform criteria, making the results more reliable. However, FoF alone cannot provide the cluster's fundamental parameters. Therefore, we use {\tt ASteCA} to obtain the cluster's fundamental parameters and, by applying the corresponding parallax restrictions, remove spuriously selected stars that do not actually belong to the Stellar Snake clusters. These steps allow us to distinguish the member stars from field stars for each cluster within the Stellar Snake.

We only select member stars in the mass range of 0.5 to 2.0 M$_\odot$ for calculating the MF to ensure that we compare results within the same selection criteria, making the comparison between clusters more reliable. We also observe that member stars at $G = 18$\,mag exhibit a broader spread in the CMD. Considering that their errors could introduce contamination from field stars that are challenging to distinguish, our restriction on the mass range of 0.5 to 2.0 M$_\odot$ helps select stars with smaller errors. However, this also results in fewer member stars per cluster which increases the uncertainties in the obtained results.

\subsection{Stellar Snake Field Stars}
We observe changing trends among the Snake clusters
To understand whether these trends are specific to the clusters or are instead characteristic of the entire filamentary Stellar Snake structure, we select several non-cluster regions (field regions) within the Snake for further investigation. 

\begin{figure}
\centering
\includegraphics[width=1.0\linewidth]{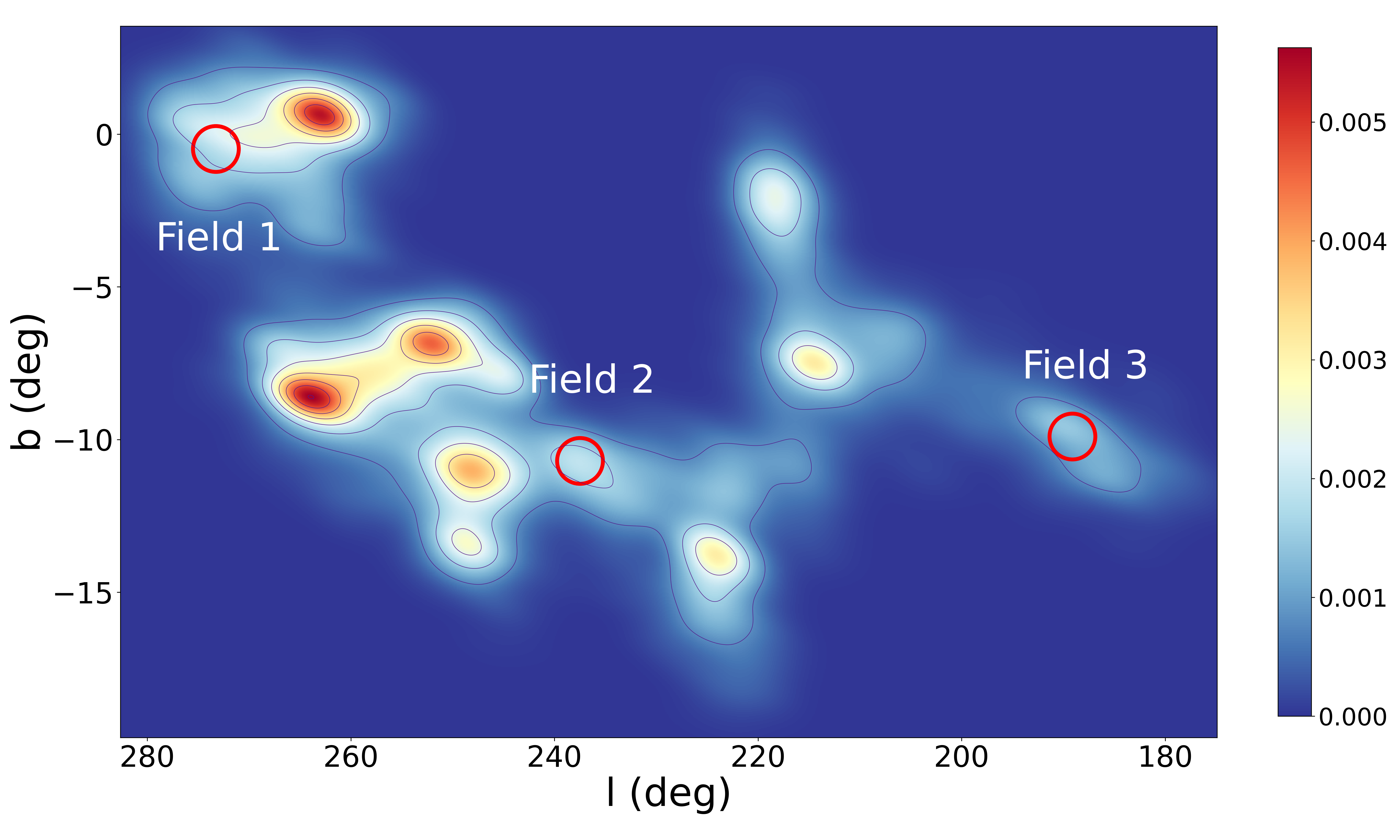}
\caption{\label{fig:Snake KDE}KDE distribution map of the Stellar Snake in the spatial coordinates $l$ and $b$. Different colors, as indicated by the colorbar, represent varying stellar number densities. The regions enclosed by the red circles represent our selected field star samples belonging to the Snake.}
\end{figure}

We initially compute the kernel density estimate (KDE) distribution map for the spatial structure of the Snake member stars, as shown in Figure \ref{fig:Snake KDE}. Next, we select three regions with a radius of 1.5 degrees from the ``head," ``middle," and ``tail" parts of the Snake as non-cluster samples (field samples Field\,1, Field\,2, and Field\,3) with the aim to calculate their MF. Considering the significant differences in distances and extinction among the member stars in non-cluster regions, we directly convert the apparent magnitudes to absolute magnitudes using 
\begin{equation}
M_{G} = G+5\times\log_{10}{\varpi}-10.
\end{equation} 
For extinction, we utilize the Galactic extinction catalog \citep{leike2020resolving} to assign extinction values to each member star.
\textbf{We use the extinction corrected $M_{G}$ and $G_{\rm BP\_RP}$ magnitudes for all stars within a specified radius in the Snake field, with DM and $A_V$ set to 0 as input parameters. We perform the fitting by using {\tt ASteCA} to obtain the best isochrone.}
Finally, we employ our simulated CMD model to fit these three regions, and present the results in Table \ref{tab:Non-cluster}. Interestingly, we observe that the MF in these three regions also seem to follow a decreasing trend from head-to-tail, similar to the trend observed in the Stellar Snake open clusters. Therefore, we can infer that this trend appears to be present throughout the entire Snake structure, rather than being specific to cluster regions.

\begin{table*}
\centering
\caption{\label{tab:Non-cluster} 
Properties for each Stellar Snake non-cluster region.}
\scalebox{0.80}{
\begin{tabular}{c|c|c|c|c|c|c|c|c}
Name   &  RA$^*$ & DEC$^*$ &  ${\bm{\log{\mathrm{Age}}}^a}$
 & ${\bm{{Z}^a}}$ & Mass range & $\alpha^b$     & $f_b^b$   & $\gamma^b$ \\\hline
 &  \multicolumn{2}{|c|}{deg} & \textbf{dex} & & $\mathrm{M}_\odot$ & & &\\\hline
Field\,1   & 141.0&$-51.0$ & $7.703\pm0.028$ & $0.0186\pm0.0016$ &0.5$-$2.0    & $-2.09^{+0.34}_{-0.35}$ & $0.25^{+0.07}_{-0.06}$ & $-0.78^{+1.08}_{-1.03}$  \\

Field\,2   & 104.5& $-26.5$ & $7.597\pm0.025$ & $0.0195\pm0.0010$ & 0.5$-$2.0    & $-2.48^{+0.40}_{-0.43}$ & $0.25^{+0.10}_{-0.08}$ & $0.64^{+1.64}_{-1.62}$  \\

Field\,3    & 82.5& 16.0 & $7.497\pm0.016$ & $0.0166\pm0.0006$ &0.5$-$2.0    & $-2.82^{+0.47}_{-0.48}$ & $0.58^{+0.08}_{-0.09}$ & $-1.84^{+0.70}_{-0.79}$  \\
\hline
\end{tabular}}
\begin{minipage}{\textwidth}
\raggedright
\textbf{Note: } $^*$Field centers.\\
\textbf{$^a$Parameters obtained using {\tt ASteCA} (Section \ref{Asteca}).}\\
$^b$Fitting results from our simulated CMD model.
\end{minipage}
\end{table*}

\section{Conclusions}\label{sec:conclusions}
We characterize the Stellar Snake's open clusters, measuring their age, metallicity, MF, binary star fraction, and binary star mass ratio distribution, by utilizing the high quality and quantity of data from $\it{Gaia}$ DR3. 

We employ the FoF algorithm {\tt ROCKSTAR} to identify a preliminary sample of candidate member stars for each Snake cluster. Subsequently, we employ {\tt ASteCA} to measure parameters such as the cluster radius $r_\mathrm{cl}$, age, and center. We then select the final sample of member stars of each Stellar Snake open cluster as Snake members within the {\tt ASteCA}-determined $r_\mathrm{cl}$. Through {\tt ASteCA}'s parameter determination, we find that the age distribution of the Snake clusters is consistent with previous expectations, falling in the range of $30-40$\,Myr. Additionally, these clusters exhibit metallicity similar to that of the Sun.

Given the high demand for completeness in determining the MF of open clusters, we conduct an assessment of data completeness within $r_\mathrm{cl}$ and the impact of our selection criteria on completeness. We observe that {\it Gaia}'s data is generally complete for $G < 18$\,mag. However, we note that limiting RUWE has a significant impact on the completeness of data within the Stellar Snake open cluster BBJ\,1. This situation may be attributed to issues intrinsic to the {\it Gaia} observations or to the presence of numerous unresolved binary stars within BBJ\,1's $r_\mathrm{cl}$. Therefore, we divide the data into two groups: one with a restriction of RUWE $< 1.4$ and another without RUWE restrictions. This division is made to compare the influence of RUWE restrictions on the results.


Subsequently, to calculate the MF ($\alpha$), binary star ratio ($f_b$), and binary mass ratio distribution ($\gamma$) for the Snake clusters, we construct a comprehensive simulated CMD model to obtain the corresponding results. After validating the reliability of the model, we compute the best-fitting results for $\alpha$, $f_b$, and $\gamma$ for each cluster within the $[0.5-2.0\, \mathrm{M}_\odot]$ mass range.
We find that the Snake cluster binary star ratios are primarily concentrated between $30-50\%$.
Concerning the MF, the comparison between the two data samples shows that, except for BBJ\,1, restricting RUWE did not lead to significant differences in the MF.

Additionally, we observe similar trends in both samples such that tracing the Snake from head-to-tail reveals a trend with
the power law index $\alpha$ of the MF becoming more negative. 
We find that the MF of the star clusters near the Snake's head are well described by a Kroupa IMF within their respective mass ranges, whereas the four relatively young star clusters near the tail (BBJ\,1, NGC\,2232, Tian\,2, Collinder\,140) 
are not as well represented by a Kroupa IMF.
These clusters with more negative $\alpha$ imply that
within the mass range of $[0.5-2.0 \,\mathrm{M}_\odot]$, clusters with more negative $\alpha$ have a relatively higher proportion of low-mass stars compared to the typical Kroupa IMF ($\alpha = -2.3$). Similarly, we also calculate the MF for the field stars in the ``head,'' ``middle,'' and ``tail'' regions of the Snake. We find the same trend is  also present in the field star sample.
One plausible explanation for this trend is that the Stellar Snake is a hierarchically primordial structure. The head of the structure (e.g., Trumpler\,10, the physical pair of Collinder\,135 and UBC\,7, NGC\,2547, and NGC\,2451B) formed initially, and around a few Myrs later, the formation of the Snake's tail (e.g., Tian\,2) commenced.
Due to the delayed formation of massive stars, when massive stars formed in clusters at the head of Snake, their feedback in the form of stellar winds blew away the surrounding gas, gradually halting star formation. The four relatively young clusters in the tail of Snake (BBJ\,1, NGC\,2232, Tian\,2, Collinder\,140), influenced by the stellar winds, did not form as massive stars and were consequently forced to cease star formation before doing so, thus resulting in the lack of massive stars.


\section*{Acknowledgements}
We thank Zhi-Yu Zhang, Chao Liu, and Feng Wang for helpful discussions and acknowledge the National Natural Science Foundation of China (NSFC) under grant Nos. 12373033, 12303023; the Cultivation Project for LAMOST Scientific Payoff and Research Achievement of CAMS-CAS; and the science research grants from the China Manned Space Project including the CSST Milky Way and Nearby Galaxies Survey on Dust and Extinction Project CMS-CSST-2021-A09 and No. CMS-CSST-2021-A08. This work has made use of data from the European Space Agency (ESA) mission {\it Gaia} (https://www.cosmos.esa.int/gaia), processed by the {\it Gaia} Data Processing and Analysis Consortium (DPAC, https://www.cosmos.esa.int/web/gaia/dpac/consortium). Funding for the DPAC has been provided by national institutions, in particular the institutions participating in the {\it Gaia} Multilateral Agreement. Guoshoujing Telescope (LAMOST) is a National Major Scientific Project built by the Chinese Academy of Sciences. Funding for the project has been provided by the National Development and Reform Commission. LAMOST is operated and managed by the National Astronomical Observatories, Chinese Academy of Sciences.

\section*{Data Availability}
The data supporting this article will be shared upon reasonable request sent to the corresponding authors.



\bibliographystyle{aasjournal}
\bibliography{sample} 




\appendix

\section{{\tt ASteCA} best-fitting parameters}

\begin{figure*}
\centering
\includegraphics[width=0.6\linewidth]{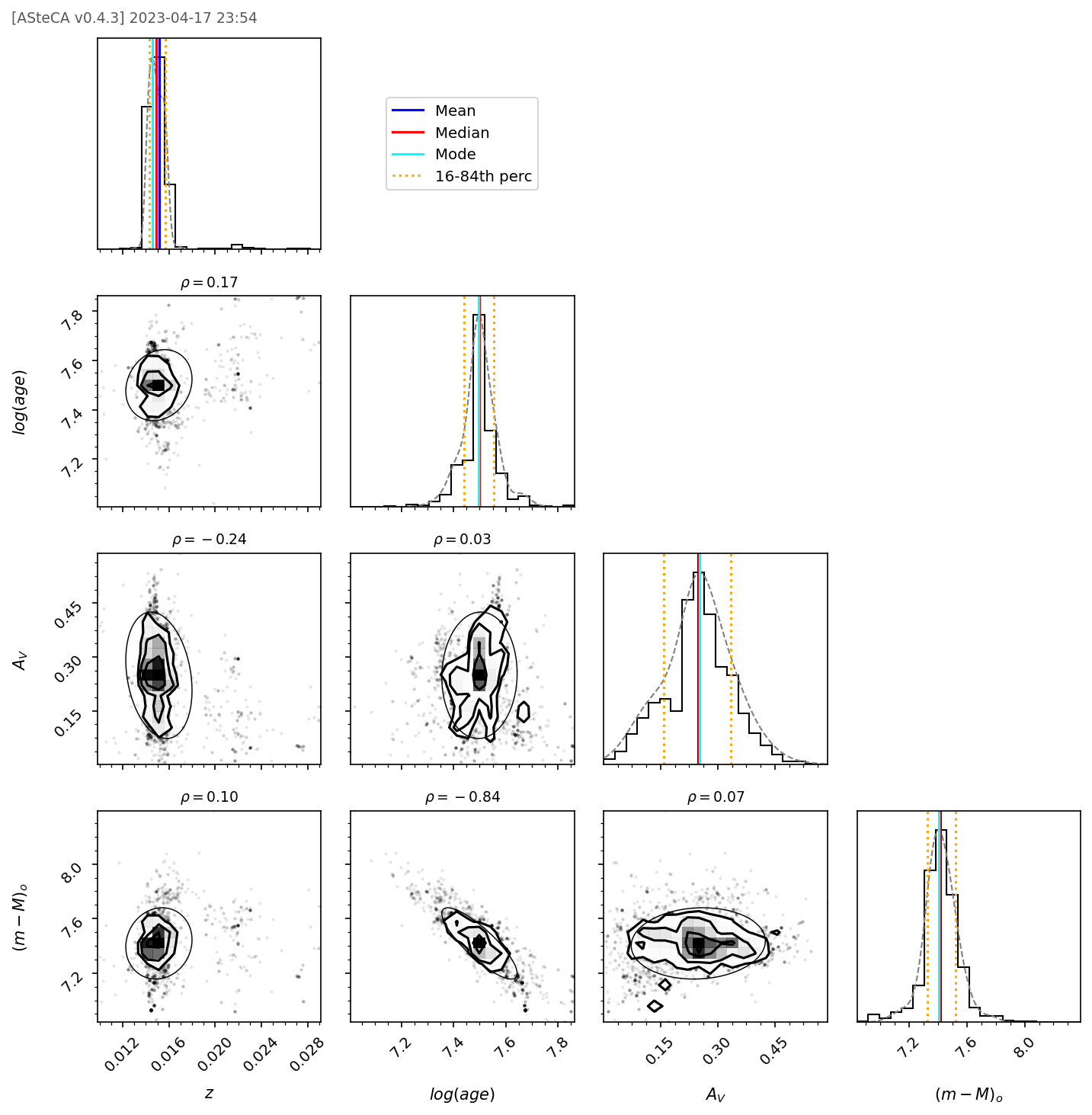}
\caption{\label{fig:NGC2232_B}
Corner plot of the MCMC-derived posterior parameters for NGC 2232 in the Stellar Snake cluster, including metallicity (Z), logarithmic age (log Age), extinction ($A_V$), and distance modulus ($\mathrm{DM}=(m-M)_0$).
}
\end{figure*}

Figure \ref{fig:NGC2232_B} presents the best-fitting open cluster parameters from {\tt ASteCA} using NGC\,2232 as a example. The black histogram represents the 1D posterior results for each parameter. For plotting purposes, we use Kernel Density Estimation (KDE) to show the 2D posteriors (contours) along with the 1D posteriors (gray dashed-line associated with the probability distribution obtained by KDE). The black thin-lined ovals of the 2D posteriors describe the correlations ($\rho$, values indicated above each panel) between different parameters. Shown with the 1D posteriors are the blue vertical line representing the fitted mean value, the red vertical line representing the fitted median value, the cyan vertical line representing the peak value of the Gaussian fit, and the yellow dashed vertical lines representing the 16th and 84th percentiles. For $\log{\mathrm{Age}}$, $A_V$, and DM, the mean, median, and mode values largely overlap.

\section{Identification of Open Clusters within the Stellar Snake}

\begin{figure*}
\centering
\includegraphics[width=1.0\linewidth]{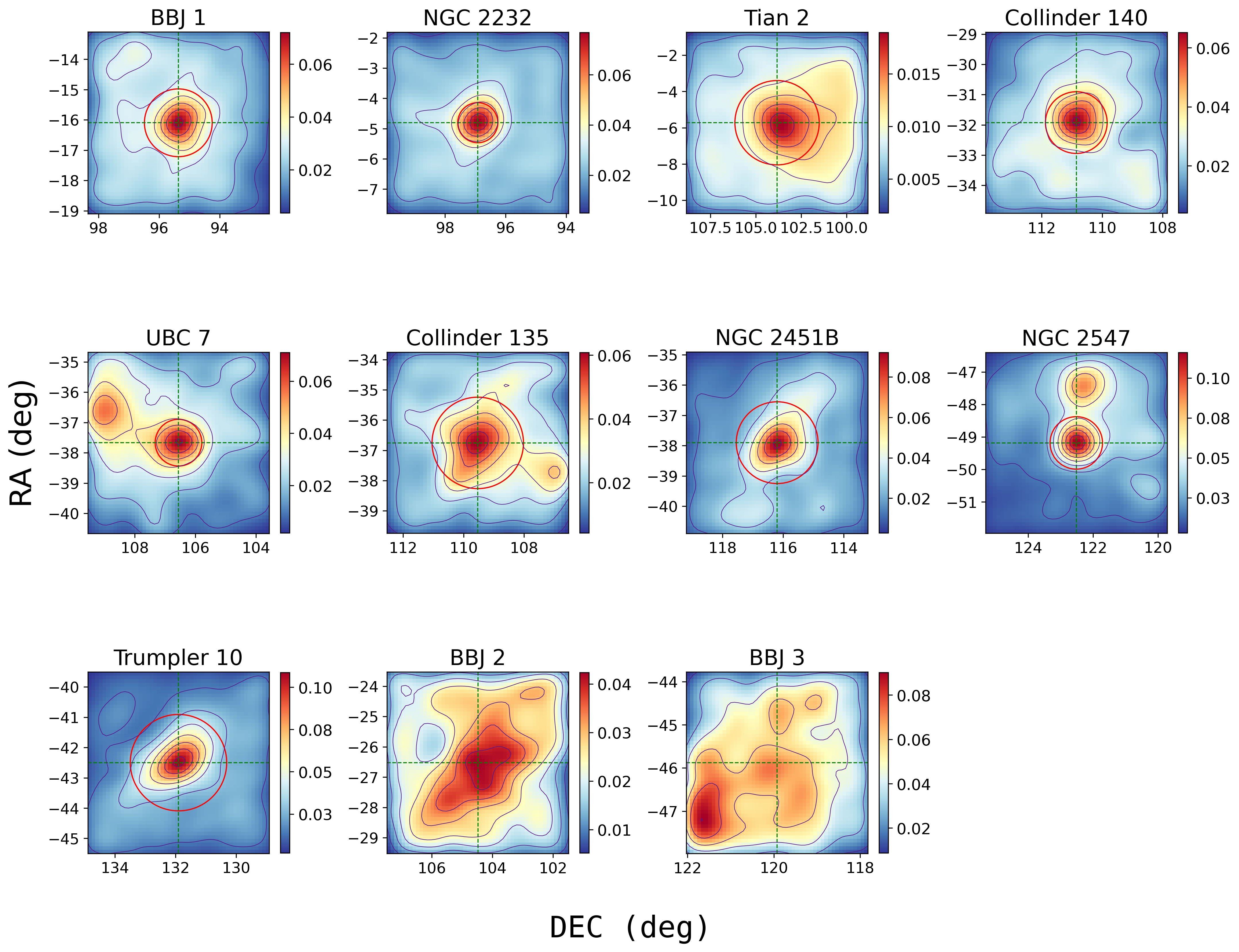}
\caption{\label{fig:KDE}Kernel density estimation plot of the central region of the open clusters. The green dashed lines intersect at the {\tt ASteCA}-determined center of the cluster, while the red circle denotes the {\tt ASteCA}-determined cluster radius $r_\mathrm{cl}$. The blue curves represent density contour lines. The different colors on the colorbars represent the varying star number density within different regions. BBJ3 and BBJ3 lack red circles because {\tt ASteCA} is unable to determine their centers.}
\end{figure*}

\citet{beccari2020uncovering} mentions the simultaneous discovery of open clusters BBJ\,2 and BBJ\,3 along with the filamentary structure in which they are hosted. These two cluster components are also part of the Snake. \citet{wang2022stellar} classify BBJ\,2 and BBJ\,3 as star clusters and provide their corresponding cluster parameters. However, upon further investigation, we find that BBJ\,2 and BBJ3 do not satisfy the qualifications of traditional open clusters. Their densities are notably lower compared to other star clusters and, as depicted in Figure \ref{fig:KDE}, their density distribution does not exhibit a clear decreasing trend from the cluster center outward, unlike other open clusters. We are unable to reliably determine the cluster radius using {\tt ASteCA}. In comparison to the other open clusters, we find that BBJ\,2 and BBJ\,3 do not show as strongly the characteristics of being open clusters. Therefore, as a precaution, we exclude BBJ\,2 and BBJ\,3 in our open cluster analysis. Instead, we promote that characterizing them as part of the main filamentary structure is more appropriate. 

\section{Evaluation of Simulated CMD Model}\label{sec:Model Validation}
\begin{figure*}
\centering
\includegraphics[width=0.6\linewidth]{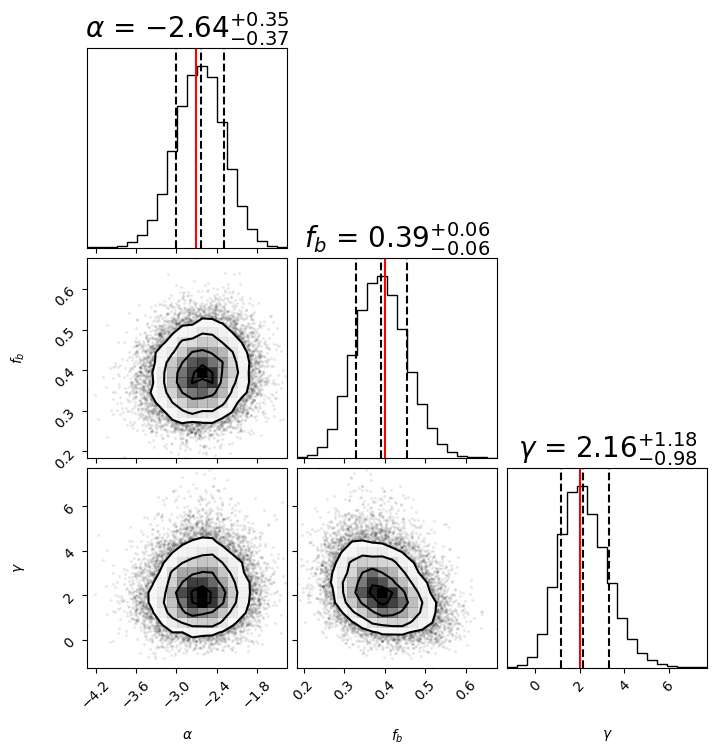}
\caption{\label{fig:MC result}
Simulated CMD model evaluation using
the {\tt emcee} 1D and 2D posterior samples.
The red vertical lines represent the true values of the parameters, while the black dashed vertical lines indicate the 18th, 50th, and 84th percentiles (also shown above each 1D posterior panel). 
The parameters being fit include $\alpha$ (MF power-law index), $f_b$ (binary fraction for $q > 0.3$) and $\gamma$ (power-law index of the distribution of binary mass ratios).}
\end{figure*}

\begin{figure}
\centering
\includegraphics[width=1.0\linewidth]{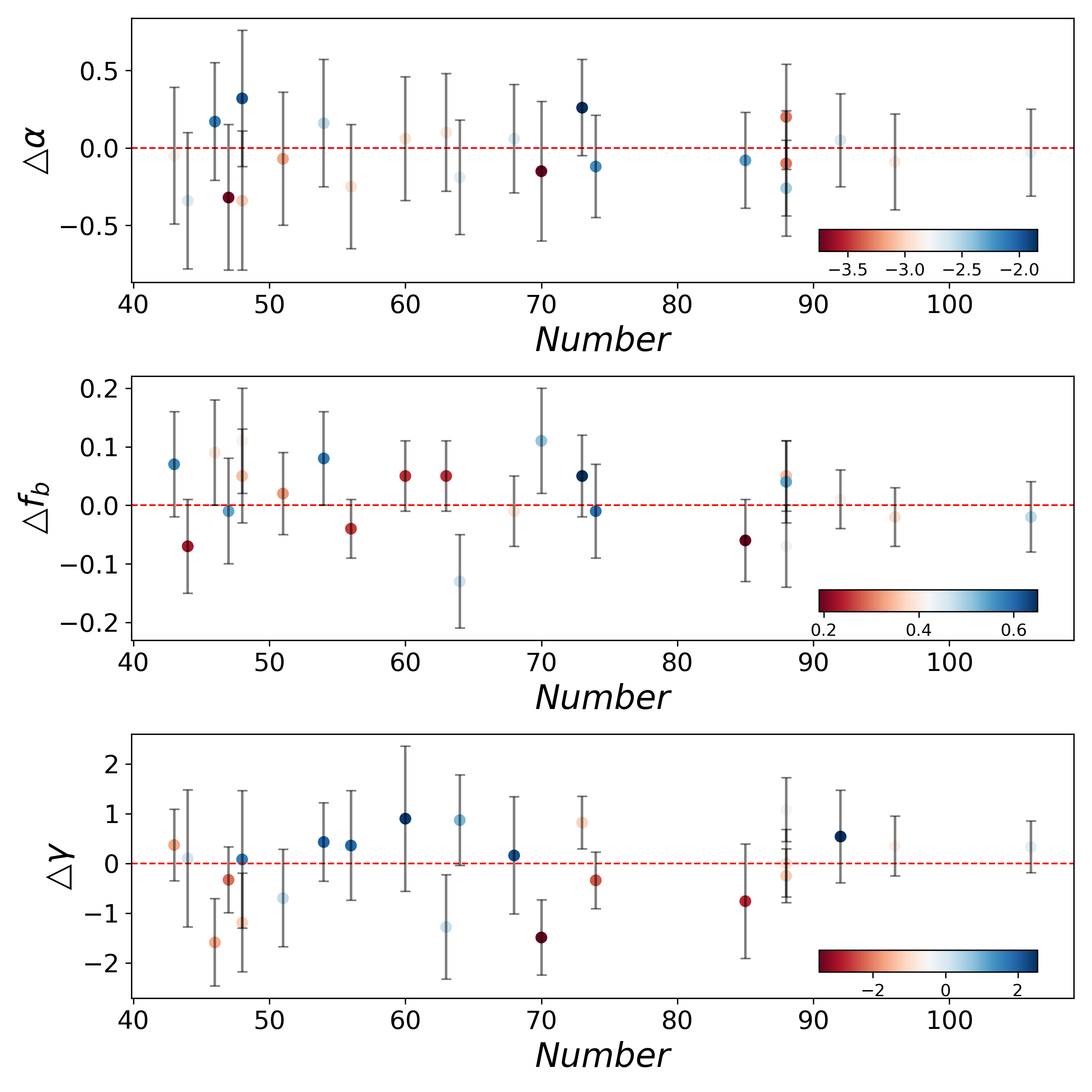}
\caption{\label{fig:Residual error}Residuals (fitting values $-$ true values) for $\alpha$, $f_b$, and $\gamma$. The horizontal axis represents the number of stars used in the fitting process, while the vertical axis represents the residuals. The error bars indicate the 1$\sigma$ range of errors on the fitting values. The different colors on the colorbar represent the various real values used for the synthetic stars. }
\end{figure}

To test the reliability of our simulated CMD model, we conduct multiple tests to verify if it can recover the true $\alpha$, $f_b$ and $\gamma$ of clusters. We follow these steps:
\begin{enumerate}
\item  We generate a random number of synthetic star clusters with a similar number of member stars as observed in the Stellar Snake star clusters, using the methods mentioned in section \ref{Sim star}. To better replicate the observational data, we use a broken power-law form to generate the mass function of the member stars. Specifically, we calculate the mass function between $[M_1, M_2]$ (where $M_1 < M_2$) by using a broken power-law with a power-law index of $\alpha$ for the mass function between [$M_1, M_2$]. We use a power-law index of $\alpha_1$ for masses lower than $M_1$ in order to generate a fraction of stars in the low-mass range. 
This approach better represents the real scenario as it takes into account the presence of unresolved binary systems where the primary star may have a lower mass than $M_1$ but, due to its near-mass equality with the secondary star, results in a brighter overall magnitude. These unresolved binary systems may be included within the fitting range, even if their primary star has a lower mass than $M_1$, potentially affecting the fitting results.

\item We use our simulated CMD model (Section \ref{Sim star} and the additional modification described in the above step [i]) to fit the mass function of the synthetic star clusters and obtain their best-fitting values. This test result is illustrated in Figure \ref{fig:MC result}.

\item We conduct multiple tests by varying the values of $\alpha$, $f_b$, and $\gamma$, as well as considering different numbers of stars for fitting. This approach allows us to obtain the residuals between the true values of the synthetic clusters and the fitted values. The relationship between the residuals and the number of stars fitted is shown in Figure \ref{fig:Residual error} to validate the reliability of our model.
\end{enumerate}

From Figure \ref{fig:Residual error}, it is evident that the fitting results are satisfactory. Regarding our primary parameter of interest, $\alpha$, we observe that even with only around 40 stars used in the fitting process, all the results fall within the 1$\sigma$ range. This indicates that our model can accurately reconstruct the corresponding $\alpha$ values. As for the binary fraction $f_b$ and the binary mass ratio distribution $\gamma$, the majority of the results also lie within the 1$\sigma$ range. Taking into account the photometric errors that may hinder a clear distinction between single and binary stars, these results are more than acceptable and demonstrate that our model can reasonably reflect the true values of $f_b$ and $\gamma$. Overall, we confidently assert that our model successfully recovers the true parameters of the star clusters, even for the case with a relatively small number of stars.

Our simulated CMD model is based on generating $10^5$ synthetic stars under specific $\alpha$, $f_b$, and $\gamma$ conditions using the best-fit PARSEC isochrones. The density distribution of these synthetic stars is obtained through KDE. By inputting cluster member star data, our model calculates the probability of generating this distribution under specific $\alpha$, $f_b$, and $\gamma$. Finally, MCMC is used to generate different $\alpha$, $f_b$, and $\gamma$ to obtain the best-fit results.

From a model validation perspective, our simulated CMD model performs reasonably well, even with a limited number of member stars. However, like other models, its accuracy relies on the accuracy of the given PARSEC isochrones and whether these isochrones accurately represent the true conditions of the star cluster. Additionally, the accuracy of the mass values provided by the isochrones can also impact the accuracy of the results. Another point to note is that our simulated CMD model generates synthetic stars with added errors based on isochrones as a reference to calculate density distributions. Consequently, for stars that deviate significantly from the isochrones, i.e., those in regions not covered by the synthetic stars, the returned probability values may be lower.

Our simulated CMD model is similar to the approach used in \cite{ebrahimi2022family}, where $10^5$ 
stars are generated based on the best-fit isochrones to model the true distribution of stars in a star cluster. However, \cite{ebrahimi2022family} uses an iterative method to obtain the final values of $\alpha$, $f_b$, and $\gamma$, while our simulated CMD model uses MCMC, which may be considered more reliable. Additionally, our simulated CMD model provides an estimate for $\gamma$ as well, although this may come at the cost of increased computation time.

The cluster modelling of \citet[][]{li2022mimo}, MIxture Model for Open clusters (MiMO), is able to skip the member star selection process and directly estimate parameters like member star fraction, best-fit isochrones, $\alpha$, $f_b$, $\gamma$, etc., when dealing with mixed populations of member and field stars. This makes {\tt MiMO} a more rigorous and comprehensive approach. In contrast, our simulated CMD model can seamlessly integrate with different clustering methods, allowing the member stars obtained through clustering to be directly applied to our simulated CMD model.

\bsp	
\label{lastpage}
\end{document}